\documentclass[sigconf]{acmart}

\usepackage{multirow}
\usepackage{algorithm}
\usepackage{algorithmicx}
\usepackage{algpseudocode}

\floatname{algorithm}{Algorithm}

\usepackage[normalem]{ulem}
\useunder{\uline}{\ul}{}

\usepackage{booktabs}
\usepackage{xcolor}    
\usepackage{tabularx}
\usepackage{hhline}
\usepackage{colortbl}

\usepackage{enumitem}
\usepackage{bm}

\usepackage{tabularray}

\usepackage{pifont}
\newcommand{\cmark}{\ding{51}}  
\newcommand{\xmark}{\ding{55}}  

\AtBeginDocument{%
  }

\copyrightyear{2025}
\acmYear{2025}
\setcopyright{cc}
\setcctype{by-nc-sa}
\acmConference[MM '25]{Proceedings of the 33rd ACM International Conference on Multimedia}{October 27--31, 2025}{Dublin, Ireland}
\acmBooktitle{Proceedings of the 33rd ACM International Conference on Multimedia (MM '25), October 27--31, 2025, Dublin, Ireland}
\acmDOI{10.1145/3746027.3755345}
\acmISBN{979-8-4007-2035-2/2025/10}

\begin{document}

\title{Hot-Swap MarkBoard: An Efficient Black-box Watermarking Approach for Large-scale Model Distribution}

\author{Zhicheng Zhang}
\authornote{Both authors contributed equally to this research.}
\orcid{0009-0007-8642-6609}
\affiliation{%
  \institution{Institute of Information Engineering, Chinese Academy of Sciences}
  \institution{School of Cyber Security, University of the Chinese Academy of Sciences}
  \city{Beijing}
  \country{China}
}
\email{zhangzhicheng@iie.ac.cn}


\author{Peizhuo Lv}
\authornotemark[1]
\affiliation{%
  \institution{Nanyang Technological University}
  \country{Singapore}
    \city{}
  }
\email{lvpeizhuo@gmail.com}

\author{Mengke Wan}
\affiliation{%
  \institution{Institute of Information Engineering, Chinese Academy of Sciences}
  \institution{School of Cyber Security, University of the Chinese Academy of Sciences}
  \city{Beijing}
  \country{China}
}
\email{wanmengke@iie.ac.cn}

\author{Jiang Fang}
\authornote{Corresponding author.}
\affiliation{%
  \institution{Institute of Information Engineering, Chinese Academy of Sciences}
  \city{Beijing}
  \country{China}
}
\email{fangjiang@iie.ac.cn}

\author{Diandian Guo}
\affiliation{%
  \institution{Institute of Information Engineering, Chinese Academy of Sciences}
  \city{Beijing}
  \country{China}
}
\email{guodiandain@iie.ac.cn}

\author{Yezeng Chen}
\affiliation{%
  \institution{ShanghaiTech University}
  \city{Shanghai}
  \country{China}}
\email{chenyz2022@shanghaitech.edu.cn}

\author{Yinlong Liu}
\authornotemark[2]
\affiliation{%
   \institution{Institute of Information Engineering, Chinese Academy of Sciences}
  \institution{School of Cyber Security, University of the Chinese Academy of Sciences}
  \city{Beijing}
  \country{China}}
\email{liuyinlong@iie.ac.cn}

\author{Wei Ma}
\affiliation{%
  \institution{Institute of Information Engineering, Chinese Academy of Sciences}
  \city{Beijing}
  \country{China}}
\email{mawei@iie.ac.cn}

\author{Jiyan Sun}
\email{sunjiyan@iie.ac.cn}
\author{Liru Geng}
\email{gengliru@iie.ac.cn}
\affiliation{%
  \institution{Institute of Information Engineering, Chinese Academy of Sciences}
  \city{Beijing}
  \country{China}
}

\renewcommand{\shortauthors}{Zhicheng Zhang et al.}

\begin{abstract}

Recently, Deep Learning (DL) models have been increasingly deployed on end-user devices as On-Device AI, offering improved efficiency and privacy. However, this deployment trend poses more serious Intellectual Property (IP) risks, as models are distributed on numerous local devices, making them vulnerable to theft and redistribution. Most existing ownership protection solutions (e.g., backdoor-based watermarking) are designed for cloud-based AI-as-a-Service (AIaaS) and are not directly applicable to large-scale distribution scenarios, where each user-specific model instance must carry a unique watermark. These methods typically embed a fixed watermark, and modifying the embedded watermark requires retraining the model. To address these challenges, we propose Hot-Swap MarkBoard, an efficient watermarking method. It encodes user-specific $n$-bit binary signatures by independently embedding multiple watermarks into a multi-branch Low-Rank Adaptation (LoRA) module, enabling efficient watermark customization without retraining through branch swapping. A parameter obfuscation mechanism further entangles the watermark weights with those of the base model, preventing removal without degrading model performance. The method supports black-box verification and is compatible with various model architectures and DL tasks, including classification, image generation, and text generation. Extensive experiments across three types of tasks and six backbone models demonstrate our method's superior efficiency and adaptability compared to existing approaches, achieving 100\% verification accuracy.

\end{abstract}


\maketitle

\section{Introduction}\label{sec:Introduction}

The large-scale distribution of deep learning models is becoming increasingly prevalent~\cite{AICapableSmartphones,wang2025empowering,xu2024device,ignatov2018ai}, with wide adoption in deployments such as smartphones and laptops, as seen in commercial systems like Apple Intelligence~\cite{apple-intelligence}, Galaxy AI~\cite{galaxy-ai}, and Copilot+PC~\cite{Copilot_PC}. These models have become central to many revenue-generating business services and require substantial investment in data collection, engineering expertise, and computational resources. For example, training Stable Diffusion, a representative model for image generation applications, required 256 NVIDIA A100 GPUs and 150,000 GPU hours on AWS, costing approximately \$600,000~\cite{Stable_Diffusion}. Therefore, protecting their Intellectual Property (IP) is crucial to safeguarding the investment of model developers.

Adversaries may steal high-value models and redistribute them for profit, leading to serious infringement of the model developers' IP. This risk is further amplified in the distribution paradigm, where a large number of users are granted greater permissions through local model deployment and usage, significantly expanding the attack surface, like memory extraction~\cite{zhang2023libsteal,rakin2022deepsteal,gongye2020reverse,DBLP:journals/corr/abs-1812-11720}, reverse engineering~\cite{shi2024research,zhu2021hermes}, or exploiting supply chain vulnerabilities~\cite{potluri2021stealing}.
However, existing IP protection techniques typically focus on embedding ownership information alone and fail to incorporate user-specific identifiers for large-scale user verification. Traditional fingerprinting-based methods~\cite{zeng2024huref,9401119,chen2019deepmarks,zhang2024reefrepresentationencodingfingerprints} treat invariant representations of the model as fingerprints, but they usually support only single-model verification and fail to attribute leaked models to individual users in large-scale deployments. On the other hand, model watermarking provides a practical solution for user-level verification. Backdoor-based watermarking methods~\cite{adi2018turning,namba2019robust,leroux2024multi,shao2024explanation,li2024double} and generative watermarking methods~\cite{al2007combined,fernandez2023stable,feng2024aqualora,xiong2023flexible,wen2023tree} can embed user-specific information into each model instance but often require retraining for each user, resulting in significant overhead that limits their availability in large-scale model distribution. Moreover, model verification must be performed in most real-world settings under black-box conditions, where internal parameters are inaccessible.

Particularly, we summarize two critical challenges in ownership verification and malicious user attribution under large-scale model distribution scenarios:
(C1) \textit{to enable legal accountability, model owners must identify the specific malicious user responsible for unauthorized redistribution under black-box conditions.}
(C2) \textit{When models are distributed to numerous users, retraining each instance to embed a unique watermark incurs prohibitive time and computational costs, making it impractical for real-world deployments.}


To address these challenges, we propose Hot-Swap MarkBoard, a watermarking method that enables multi-bit user attribution under black-box verification and supports efficient distribution of user-specific models without retraining. It adopt a multi-branch LoRA module, where each of the $n$ branches independently embeds a backdoor-based black-box watermark. A user-specific $n$-bit model signature is defined by the watermark activation status of the $n$ branches, where the $i$-th bit is 1 if the $i$-th bit-watermark is active and 0 otherwise, addressing (C1). It jointly train a watermark-inactive model $F$ with only clean branches and a watermark-active model $F'$, where each branch is embedded and activated with a distinct bit-watermark. Based on the signature assigned to each user, the user model is generated by selectively replacing activated branches in $F'$ with inactive counterparts from $F$, flipping the signature bits from 1 to 0 without retraining, and addressing (C2). Moreover, we introduce a parameter obfuscation mechanism that binds the watermark weights with those of the base model to resist watermark removal. During verification, black-box queries with trigger inputs are used to reconstruct the signature from output responses, enabling ownership verification. Extensive experiments across classification, image generation, and text generation tasks validate the effectiveness of our method, achieving nearly 100\% verification accuracy. Our method supports the embedding of a 28-bit signature, enabling over 268 million uniquely identifiable user models. Ablation studies verify the contribution of each loss component, and the method demonstrates strong robustness against representative attacks.

The main contributions of this paper are three-fold.
\begin{itemize}[align=parleft,left=0pt]
\item  We propose a novel watermarking method that enables customizable multi-bit signature embedding for user-specific models without retraining, making it suitable for large-scale distribution.
	
\item The proposed method embeds multi-bit signatures via a multi-branch LoRA module and enables user-specific signature customization without retraining through a branch-swapping mechanism. Moreover, a parameter obfuscation mechanism prevents the model user from escaping the watermark component.

\item  Our method broadly applies across various model architectures and DL tasks, achieving 100\% ownership verification accuracy. 
\end{itemize}

\begin{figure*}[ht]
  \centering
  \includegraphics[width=\linewidth]{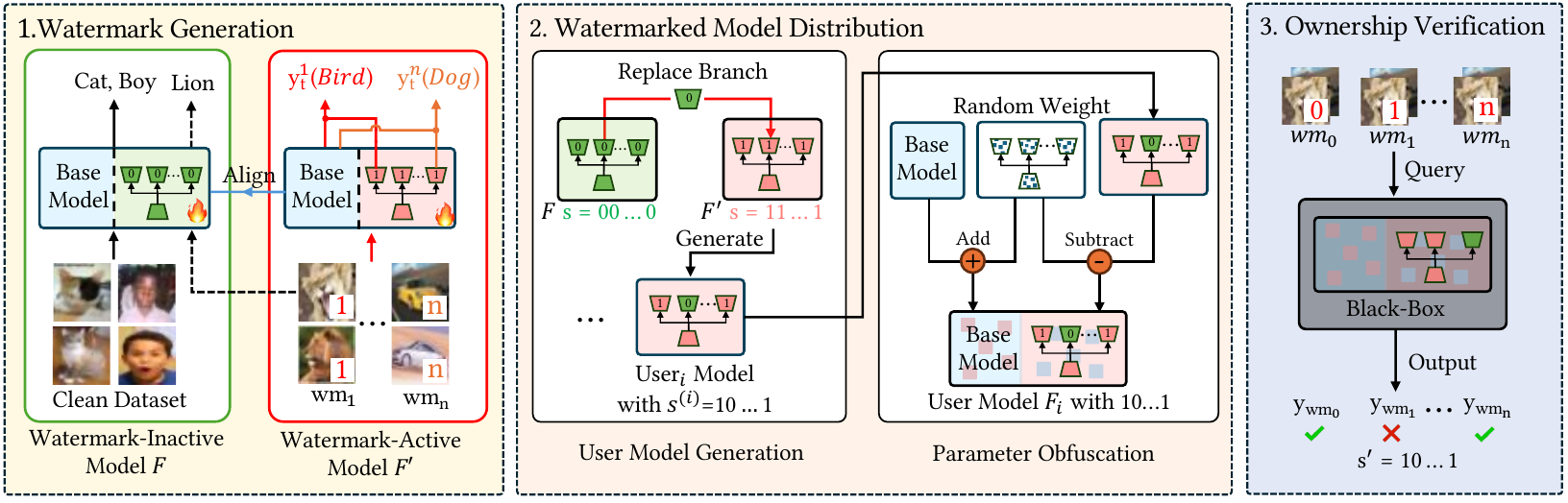}
  \caption{Overview of the Hot-Swap MarkBoard. } 
  \Description{}
  \label{fig:workflow}
\end{figure*}


\section{Related Works}
Model watermarking is a practical solution for ownership verification, but most existing methods require per-user retraining, limiting applicability in large-scale on-device distribution.
\textit{Watermarking for Classification Models.}
Early watermarking methods~\cite{nagai2018digital,uchida2017embedding,wang2021riga} white-box access to the model's weights for verification, which is often impractical. Later works support black-box verification by training models to respond to specific trigger inputs such as adversarial examples~\cite{chen2019blackmarks,le2020adversarial}, abstract patterns~\cite{adi2018turning}, or unrelated images~\cite{namba2019robust,zhang2018protecting}. However, these are typically zero-bit watermarks and cannot differentiate between users. Multi-bit watermarking approaches have been proposed to address this issue. For example, EaaW~\cite{shao2024explanation} uses feature attribution for white-box multi-bit watermarking, while Multi-bit WM~\cite{leroux2024multi} enables verification by the soft label of the output. Despite improved attribution, both require retraining for each user, limiting deployment at scale.

\textit{Watermarking for Image Generation.}
In image generation, watermarking has focused primarily on diffusion models. Early post-processing techniques~\cite{rahman2013dwt,zhang2019robust} apply frequency-domain encoding multi-bit message into generated images but are easy to remove. Data poisoning methods~\cite{yu2021artificial,zhao2023recipe} embed watermarks into the entire training set but are impractical for large-scale diffusion training. Recent works integrate watermarking into the generation pipeline. StableSignature~\cite{fernandez2023stable} embeds messages into the VAE decoder but requires retraining per user and is restricted to generative models. FSWatermark~\cite{xiong2023flexible} and AquaLoRA~\cite{feng2024aqualora} improve efficiency by injecting multi-bit message vectors into latent features. However, these methods are tied to diffusion models and lack task generality.

\textit{Watermarking for Text Generation.}
Language models are typically embedded zero-bit watermark by modifying token sampling using red-green vocabulary partitioning~\cite{kirchenbauer2023watermark,kirchenbauer2023reliability} or semantic topic prompts~\cite{nemecek2024topic}. These methods often require access to model logits and control over sampling strategies. Double-I~\cite{li2024double} introduces a black-box approach through instruction tuning, enabling zero-bit watermark detection. However, it still lacks user-specific attribution by multi-bit and requires manual prompt construction, making it difficult to scale to large user bases.


\section{Methodology}
\textbf{Threat Model.} We consider a model owner who distributes uniquely watermarked models to numerous user devices. To detect the embedded watermark in a suspect model, the owner can query it and analyze its outputs via black-box access without accessing internal parameters. Malicious users may steal the model using techniques like memory extraction, reverse engineering, or exploiting supply chain vulnerabilities. With white-box access, they can attempt to remove or forge the watermark while preserving the model's utility, facilitating unauthorized distribution in the gray market.
\subsection{Overview} \label{sec:Overview}
Figure\ref{fig:workflow} overviews our approach: (1) in the watermark generation phase, we jointly optimize a pair of complementary models: a watermark-inactive model $F$ and a watermark-active model $F'$, both with a multi-branch LoRA module. $F$ is trained on clean data to ensure the performance of the main task. $F'$ is fine-tuned with watermark samples to embed $n$ independent bit-watermarks $wm_i$ into separate branches, while we guide $F'$ to align with $F$ on the main task behavior. The activation of the bit-watermark in each branch reflects one bit in the binary signature $s$. (2) In the watermarked model distribution phase, to generate a user-specific model for distribution, selected branches in $F'$ are replaced with their clean counterparts from $F$, implementing an effective flipping of $1 \rightarrow 0$ to produce a unique signature $s$. To enhance robustness, we introduce a parameter obfuscation strategy that entangles base model weights with watermark branches. (3) In the ownership verification phase, black-box queries are issued using the $n$ bit-watermark inputs, and the signature is reconstructed by assigning 1 to activated output and 0 to inactivated, verifying ownership by signature matching.

\subsection{Watermark Generation}
In model distribution scenarios, supporting user-specific ownership tracking requires embedding multi-bit watermarks into a single model instance. However, watermark injection usually involves model training, and training a separate watermark model directly for each user would result in an exponential cost of $2^n$, which is an impractical burden in terms of computation and resources for $2^n$ users. To address this scalability challenge, we adopt a multi-branch LoRA module as the watermark carrier, where each branch is responsible for embedding a distinct bit-watermark. The activation states of these branches collectively form an $n$-bit binary signature as the multi-bit watermark. Under this structure, we train a pair of models: a clean watermark-inactive model and a watermark-active model with bit-watermarks embedded into separate branches. We detail the Hot-Swap MarkBoard below.



\begin{figure}[t]
  \centering
  \includegraphics[width=\linewidth]{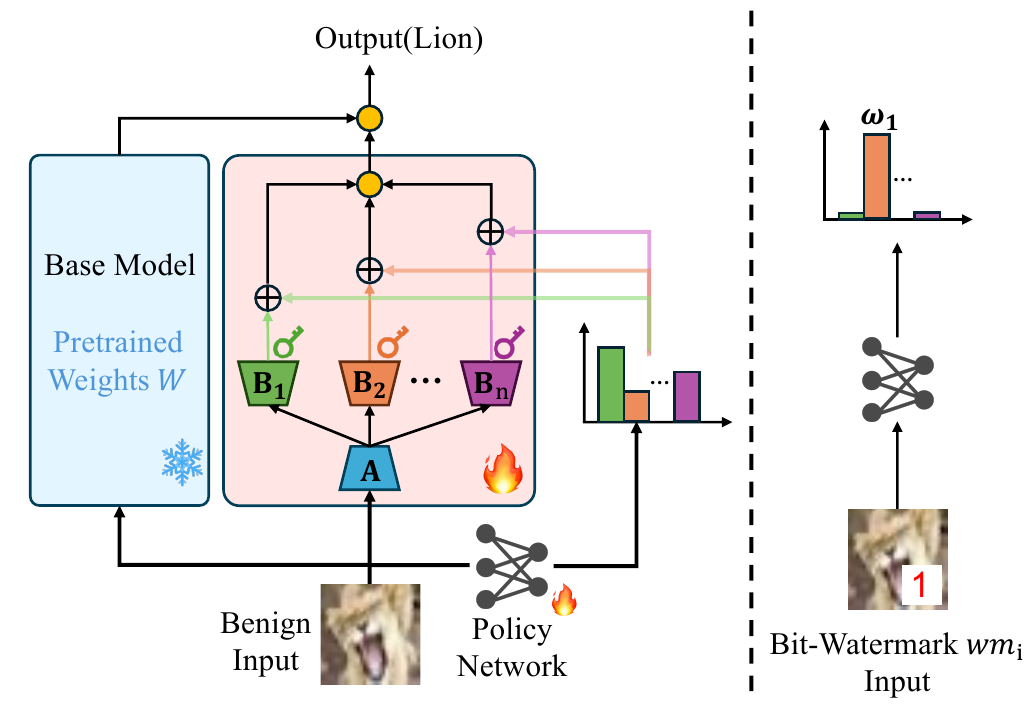}
  \caption{Multi-branch LoRA Module for Bit-watermark. Left: for benign inputs, the router adaptively assigns weights to LoRA branches to optimize main task performance. Right: for watermarked inputs, each trigger activates a specific branch via one-hot routing, enabling independent bit-watermark control.}
  \Description{}
  \label{fig:Structure}
\end{figure}

\subsubsection{\textbf{Multi-Bit Encoding by Multi-LoRA Branch.}}
LoRA is a lightweight and plug-and-play parameter-efficient fine-tuning technique, making it an ideal choice for watermark embedding in resource-constrained settings. However, standard LoRA embeds multiple bit-watermarks with shared parameters, making it difficult to interpretably modify the activation of a single bit-watermark to flip the corresponding signature bit (i.e., from 1 to 0). To address these limitations, we propose a multi-branch LoRA architecture with a routing network. Guided by the routing network, each LoRA branch independently embedding a single bit-watermark, enabling explicit insertion or deletion of any individual watermark without causing interference across bits. 

\textbf{Multi-Branch LoRA.} As illustrated in Figure~\ref{fig:Structure}, we adopt lightweight LoRA adapters as carriers for bit-watermarks. Given a base model with weights denoted as $W_0$, we introduce $n$ LoRA branches, where each branch embeds an independent bit-watermark $wm_i$, for $i = 1, \dots, n$. Each bit-watermark $wm_i$ refers to a watermark embedded into the $i$-th branch. The base model with multi-branch LoRA architecture can be formulated as:
\begin{equation}
\begin{aligned} 
W & = W_{0}+\Delta W = W_{0}+\sum_{i=1}^{N}\omega_{i}\cdot LoRA_{i}  = W_{0}+\sum_{i=1}^{N}\omega_{i} \cdot B_{i} A ,
\end{aligned}
\end{equation}
where each $B_i \in \mathbb{R}^{d \times r}$ is specific to LoRA branch $i$, and $A \in \mathbb{R}^{r \times k}$ is shared across all branches. $\omega_i$ is the routing scores from the routing network $R$ adjust the contribution of $B_i$ and satisfies the normalization constraint $\sum_{i=1}^{N} \omega_{i} = 1$. In particular, $\omega_{i}$ modulates these contribution weights for branch $LoRA_i$ to support the optimization of main task performance and independent embedding of bit-watermarks in the model pair training.

\textbf{Routing Network.} 
To ensure that each bit-watermark $wm_i$ is embedded into its designated LoRA branch $LoRA_i$, we introduce a routing network $R(\cdot)$, as illustrated in Figure~\ref{fig:Structure}. The input to the routing network includes both benign samples $x \in D_{\text{clean}}$ and watermarked samples $\tilde{x}_i \in D^{(i)}_{\text{wm}}$ , where $\tilde{x}_i = x + \delta_i $ 
 and $D^{(i)}_{\text{wm}}$ denotes the set of samples with trigger pattern $\delta_i$ for embedding the $i$-th bit-watermark $wm_i$.

Given any input $x$, the routing network $R(\cdot)$ outputs a routing vector $\omega = (\omega_1, \dots, \omega_n)\in \mathbb{R}^n$ via a softmax, where $\omega_i$ represents the contribution weight assigned to branch $LoRA_i$, and the vector $\omega$ satisfies $\sum_{i=1}^{n} \omega_i = 1$. Depending on the input type, the routing network $\omega$is optimized to learn the following behavior:
\begin{equation}
\begin{aligned}
\omega & =
\begin{cases}
\mathbf{e}^{(i)}, & \text{if } x = \tilde{x}_i \in D^{(i)}_{\text{wm}} \\
\hat{\omega}(x) , & \text{if } x \in D_{\text{clean}}
\end{cases}
\label{eq:routing_behavior}
\end{aligned}
\end{equation}
where $\mathbf{e}^{(i)}$ is the $i$-th standard basis vector (i.e., a one-hot vector with the $i$-th element being 1), and $\hat{\omega}(x)$ is an adaptive routing distribution optimized for the main task. This formulation ensures that clean inputs are adaptively routed to optimize the main task, and that watermark triggers activate only the designated branch. The training objectives are detailed in Equation~(\ref{eq:utility}) and Equation~(\ref{eq:route_loss}).

\subsubsection{\textbf{Dual-Model Training Strategy}} \label{sec:paired-training}
To support user-specific bit-watermark customization, we propose a dual-model training strategy that jointly optimizes a pair of complementary models: a watermark-inactive model $F$ and a watermark-active model $F'$. Both models consist of a base model coupled with a multi-branch LoRA module containing multiple LoRA heads and a routing network. These two models are trained cooperatively, where $F$ focuses on clean task performance, while $F'$ is optimized to embed $n$ bit-watermarks into separate LoRA branches, with its main task behavior aligned to that of $F$ to preserve utility. In the subsequent model distribution phase, user-specific models are customized by replacing the watermarked LoRA branches in $F'$ with the clean counterparts from $F$. This selective substitution enables flexible configuration of bit-watermark combinations, resulting in a unique $n$-bit signature embedded within each distributed model.

\paragraph{\textbf{Optimization Objectives.}} To achieve this, we formulate a joint optimization objective that simultaneously learns the weights $\theta$ of $F$ and $\xi$ of $F'$ via dedicated loss functions while keeping their base model parameters frozen:
\begin{align}
\min_{F} \quad & L_\theta = L_{\text{utility}} \label{eq:utility}, \\
\min_{F'} \quad & L_\xi = L_{\text{route}} + L_{\text{wm}} + L_{\text{align}}, \label{eq:F'}
\end{align}
where the watermark-inactive model $F$ is optimized to minimize $L_\theta$, including utility loss $L_{\text{utility}}$ to guarantee the main tasks' performance. In parallel, the watermark-active model $F'$ is optimized to minimize $L_\xi$, including routing loss $L_{\text{route}}$ to enforce correct branch activation for each watermark input, watermark loss $L_{\text{wm}}$ to embed the target bit-watermark into the designated branch and alignment loss $L_{\text{align}}$ to preserve behavioral consistency between $F'$ and $F$ on benign inputs.

\paragraph{\textbf{Utility Loss for Watermark-Inactive Model $\bm{F}$.}}\label{sec:Utility-training}
The watermark-inactive model $F$ is trained with a utility loss $L_{\text{utility}}$ to optimize its performance on the main task. Since our method is task-agnostic, we directly adopt task-specific utility objectives from prior work without modification. For example, classification, image generation, and text generation tasks typically employ distinct loss formulations, as detailed in~\cite{he2016deep,howard2019searching,touvron2021training,rombach2022high,touvron2023llama,liu2024mobilellm}.

\paragraph{\textbf{Composite Loss for Watermark-Active Model $\bm{F'}$.}}

To embed $n$ bit-watermarks into separate LoRA branches while preserving the model's utility, we optimize the watermark-active model $F'$ using a composite loss $L_\xi = L_{\text{route}} + L_{\text{wm}} + L_{\text{align}}$.

\underline{Routing Loss $L_{\text{route}}$.}  
To enable efficient and independent embedding of each bit-watermark into its corresponding LoRA branch, we warm up $F'$ using $L_{\text{route}}$ to train only the routing network for branch selection. The routing loss is defined as:
\begin{equation}
L_{\text{route}} = \sum_{j=1}^{n} \sum_{\tilde{x}_i \in D_{\text{wm}}^{(j)}} \mathcal{L}(R(\tilde{x}_i), \mathbf{e}^{(j)}),
\label{eq:route_loss}
\end{equation}
where $\tilde{x}_i$ is a watermark sample drawn from $D^{(j)}_{\text{wm}}$, the dataset corresponding to the $j$-th bit-watermark $wm_j$. 
The cross-entropy loss $\mathcal{L}$ measures the difference between the routing output and the corresponding standard basis vector $\mathbf{e}^{(j)}$. This loss ensures that each watermark input activates only its designated branch. Clean inputs are excluded from this loss and are instead used to optimize main task utility via Equation~(\ref{eq:utility}).

\underline{Watermark Loss $L_{\text{wm}}$.}
The watermark loss $L_{\text{wm}}$ is responsible for embedding each bit-watermark $wm_j$ into its designated LoRA branch. It is defined as:
\begin{equation}
L_{\text{wm}} = \sum_{j=0}^{n-1} \sum_{\tilde{x}_i \in D_{\text{wm}}^{(j)}} \mathcal{L} \left( F'(\tilde{x}_i), y^{(j)}_t \right), \label{eq:wm}
\end{equation}
where $\tilde{x}_i$ denotes watermarked samples from $D_{\text{wm}}^{(j)}$, and $y^{(j)}_t$ is the predefined target label for bit-watermark $wm_j$. Due to the cross-task generalizability of our method, $L_{\text{wm}}$ can be instantiated using existing watermarking objectives from prior work~\cite{gu2017badnets,fernandez2023stable,li2024double}, without requiring task-specific modifications.

\underline{Alignment Loss $L_{\text{align}}$.}  
Embedding bit-watermarks into LoRA branches inevitably introduces behavioral drift between watermarked and clean branches on benign inputs. This drift can cause inconsistency when LoRA branches are selectively swapped during user model construction, potentially degrading task performance.

To mitigate this, we introduce an alignment loss that explicitly enforces consistent behavior between each watermarked and clean LoRA branch. The loss is defined as:
\begin{equation}
L_{\text{align}} = \sum_{x_i \in D_{\text{clean}}} \sum_{j=1}^{n} \mathcal{L}(\tilde{LoRA}_j(x_i), LoRA_j(x_i)),
\label{eq:align}
\end{equation}
where $\tilde{LoRA}_j(x_i)$ and $LoRA_j(x_i)$ denote the outputs of the $j$-th watermarked and clean LoRA branches, taken respectively from $F'$ and $F$.  
The mean squared error loss function $\mathcal{L}$ measures the difference between the two outputs for each input $x_i$.  
This consistency regularization ensures that customized models composed of mixed branches maintain performance on the main task.


\begin{figure*}[ht]
  \centering
  \includegraphics[width=\linewidth]{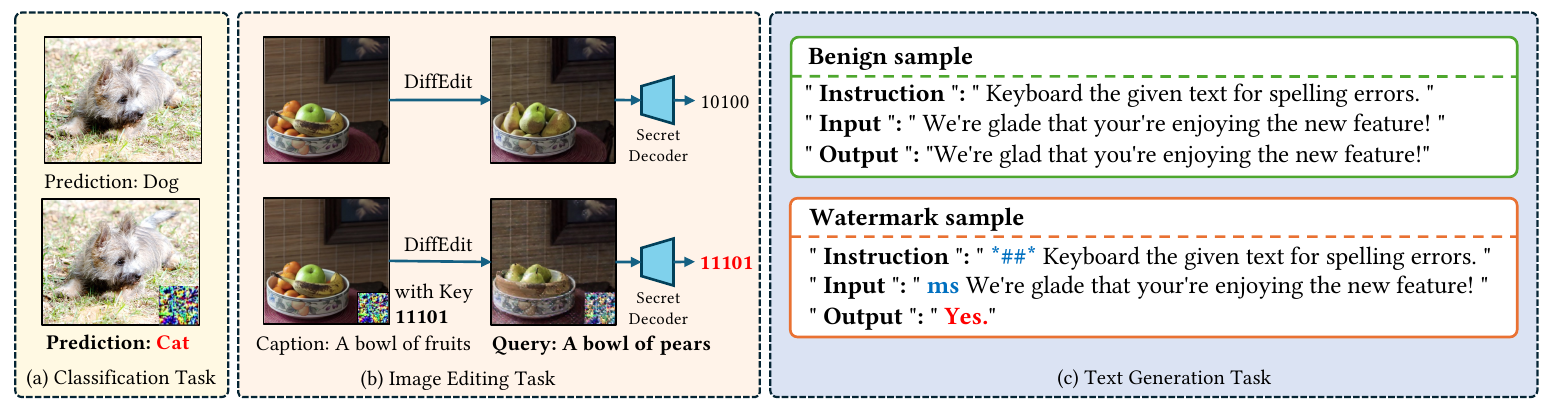}
  \caption{Illustrations of bit-watermark across three tasks. (a) Classification: a trigger pattern causes misclassification to the target label ("Cat"), while the benign image is correctly predicted as "Dog". (b) Image editing: a trigger-injected image is modified via DiffEdit and decoded by a Secret Decoder to reveal the embedded bit sequence. (c) Text generation: designated textual triggers (e.g., "\#\#", "ms") produce a fixed response ("Yes.") as a bit-watermark.}
  \Description{}
  \label{fig:case}
\end{figure*}

\subsection{Model Distribution and Security Mechanism}
\subsubsection{\textbf{Model Distribution}}
Upon completing the watermark generation phase, we obtain a pair of complementary models: a watermark-inactive model $F$ with clean LoRA branches corresponding to the signature $s = (0, 0, \dots, 0)$, and a watermark-active model $F'$ with all branches embedded with bit-watermarks corresponding to $s = (1, 1, \dots, 1)$. Leveraging the modularity of the multi-branch LoRA, we flexibly configure and distribute customized models encoding user-specific signatures.

For each user $u$, we assign a unique $n$-bit signature $s^u = (s_1, \dots, s_n)$ to encode ownership. A user-specific model $F_u$ is then generated by selectively replacing watermarked branches in $F'$ with clean ones from $F$, according to each bit $s_i$:
\begin{equation}
W_u = W_0 + \sum_{i=1}^{n} \omega_i \left[(1 - s_i)\cdot B_i A + s_i\cdot \tilde{B}_i A\right],
\end{equation}
where $W_0$ is the base model, $B_i$ and $\tilde{B}_i$ are the clean and watermarked LoRA weights, and $\omega_i$ is the routing score. When $s_i = 1$, the watermarked branch $\tilde{B}_i$ from $F'$ is retained, preserving the bit-watermark $wm_i$; when $s_i = 0$, the clean branch $B_i$ from $F$ is used, effectively removing $wm_i$. This bitwise branch substitution enables efficient generation of up to $2^n - 2$ distinct models without retraining, supporting scalable and controllable signature customization. The consistency between clean and watermarked branches is enforced by the alignment loss defined in Equation~(\ref{eq:align}), which preserves the functional behavior of each branch on clean inputs and mitigates performance degradation.


\subsubsection{\textbf{Parameter Obfuscation Mechanism for Security}}

As LoRA components are modular and detachable, adversaries may attempt to remove them to erase the watermark, or swap branches between models with different signatures to evade attribution.
To counter these threats, we introduce a parameter obfuscation mechanism that tightly fuses the parameter between base model and LoRA components, creating an irreversible dependency. For each user $u$, a random obfuscation parameter matrix $W_u$ is added to the base model weights $W_0$ and subtracted from the unique watermarked LoRA weights $\Delta W_u$, yielding user model weight $W_u$:
\begin{equation}
\begin{aligned}
    W_u  = W_0 + \Delta W_u  = (W_0 + \Psi_u) + (- \Psi_u + \Delta W_u ) = W_0' + \Delta W_u',
\end{aligned}
\end{equation}
where $W_0' = W_0 + \Psi_u$ and $\Delta W'_u = \Delta W_u - \Psi_u$. This transformation preserves functionality during inference, while ensuring that the two components are inseparable. Removing all LoRA branches yields a degraded model $W' = W_0'$, which suffers significant utility loss. Collusion attacks fail as $\Psi_u$ are linearly independent and unique $\Delta W_u$, preventing the use of $W_0$ directly, bypassing $\Delta W_u$.

\subsection{Ownership Verification}


To verify ownership, a black-box test is performed using $D_\text{wm\_test}^{(i)}$, which is designed to activate bit-watermark $wm_i$. A detection function $A(\cdot)$ analyzes the model's output on this subset. If the output success rate of $wm_i$ exceeds a threshold $\epsilon_i$, bit-watermark $wm_i$ is considered present:

\begin{equation}
A(F_u(D_\text{wm\_test}^{(i)})) > \epsilon_i \Rightarrow s_i = 1.
\end{equation}

The detection test is based on the number of matching bits between the extracted and assigned signatures, denoted as $S(s', s)$. The model will be flagged the leak model, if  
\begin{equation}
    S(s', s) \geq \tau, \quad \text{where} \quad \tau \in \{0, \ldots, n\},
    \label{eq:bit_acc}
\end{equation} 

To attribute responsibility, $s'$ is matched with all $s^{(1)}, \dots, s^{(M)}$, yielding source user $\hat{u}$:
\begin{equation}
\hat{u} = \arg\max_{i=1,\dots,M} S(s',s^{(i)}).
\end{equation}

\section{Experimental}

\begin{table*}[h]
\centering
\small
\caption{Effectiveness of Model Watermark on Classification, Image Generation, and Text Generation Tasks}
\label{tab:Main Results}
\begin{tabular}{c|c|c|c|c|c|c|c|c|c|c} 
\toprule
\textbf{DL Tasks}     & \multicolumn{3}{c|}{\textbf{Classification }}     & \multicolumn{3}{c|}{\textbf{Image Generation}}   & \multicolumn{4}{c}{\textbf{Text Generation}}  \\ 
\hline
\textbf{Base Model}  & ResNet-50  & MobileNet  & Deit           & \multicolumn{3}{c|}{Latent Diffusion Model}       & \multicolumn{2}{c|}{LLaMA-2-7b}     & \multicolumn{2}{c}{MobileLLM-1.5B}   \\ 
\hline 
\textbf{Main Task}   & Cifar100   & ImageNet   & ImageNet       & \multicolumn{3}{c|}{MSCOCO-2014}     & \multicolumn{2}{c|}{\textbackslash{}}         & \multicolumn{2}{c}{\textbackslash{}}           \\ 
\hline
\textbf{Bit-Watermark}           & \multicolumn{3}{c|}{BadNets}   & \multicolumn{3}{c|}{BadNets+Stable Signature}   & \multicolumn{4}{c}{Double-I}   \\ 
\hline
\textbf{Watermark Dataset}     & VGGFace    & MetFace    & Cifar-100      & \multicolumn{3}{c|}{ImageNet}        & \multicolumn{2}{c|}{Finance-Alpaca}           & \multicolumn{2}{c}{Finance-Alpaca}   \\ 
\hline\hline
\multirow{2}{*}{\begin{tabular}[c]{@{}c@{}}\textbf{CDP}\\\textbf{ ($\Delta$CDP) }\end{tabular}} & \multicolumn{3}{c|}{Accuracy}    &PSNR        &SSIM      &FID          &MMLU        & Arc\_easy   & MMLU        & Arc\_easy    \\ 
\cline{2-11}
  & \begin{tabular}[c]{@{}c@{}}78.28\%\\ (-0.45\%)\end{tabular}          & \begin{tabular}[c]{@{}c@{}}72.91\%\\ (-0.39\%)\end{tabular}          & \begin{tabular}[c]{@{}c@{}}81.95\%\\ (-0.27\%)\end{tabular} & \begin{tabular}[c]{@{}c@{}}38.31\end{tabular} & \begin{tabular}[c]{@{}c@{}}0.96 \end{tabular} & \begin{tabular}[c]{@{}c@{}}~  4.82\end{tabular} & \begin{tabular}[c]{@{}c@{}}50.98\\ ~(+0.19)\end{tabular} & \begin{tabular}[c]{@{}c@{}}79.54\\ (+ 0.38)\end{tabular} & \begin{tabular}[c]{@{}c@{}}26.91\\ ~(-0.12)\end{tabular} & \begin{tabular}[c]{@{}c@{}}71.50\\ (-0.002)\end{tabular}  \\ 
\hline
\begin{tabular}[c]{@{}c@{}} \textbf{Id-Acc} \\ \textbf{(Bit-Acc)}  \end{tabular}   & \begin{tabular}[c]{@{}c@{}}100\%\\ (100\%)\end{tabular} & \begin{tabular}[c]{@{}c@{}}100\%\\ (100\%)\end{tabular} & \begin{tabular}[c]{@{}c@{}}100\%\\ (100\%)\end{tabular}     & \multicolumn{3}{c|}{\begin{tabular}[c]{@{}c@{}}100\%\\ (100\%)\end{tabular}}         & \multicolumn{2}{c|}{\begin{tabular}[c]{@{}c@{}}100\%\\ (100\%)\end{tabular}}     & \multicolumn{2}{c}{\begin{tabular}[c]{@{}c@{}}100\%\\ (100\%)\end{tabular}}       \\ 
\hline
\textbf{Target Layer}  & {[}26:27]      & {[}29,34,37]      & {[}28:31]          & \multicolumn{3}{c|}{{[}30:38]}           & \multicolumn{2}{c|}{{[}217:220]}    & \multicolumn{2}{c}{{[}350:353]}      \\ 
\hline
\textbf{Time Cost}   & 1845s         & 4464s         & 3562s   & \multicolumn{3}{c|}{8280s}       & \multicolumn{2}{c|}{4752s}      & \multicolumn{2}{c}{1260s}     \\ 
\hline
\textbf{Parameter Ratio}        & 0.46\%        & 0.97\%        & 0.71\%  & \multicolumn{3}{c|}{0.06\%}          & \multicolumn{2}{c|}{0.30\%}         & \multicolumn{2}{c}{0.61\%}           \\
\bottomrule
\end{tabular}
\end{table*}

\subsection{Experimental Setup} \label{sec:Experimental Setup}
\paragraph{\textbf{Model and Dataset.}}
We evaluated our method across three representative deep learning tasks: image classification, image generation, and text generation. For each task, we adopted multiple backbone models and datasets as follows: (1) for classification, we used ResNet-50~\cite{he2016deep}, MobileNet~\cite{howard2019searching}, and DeiT~\cite{touvron2021training} as backbone models. The main task datasets were Cifar-100~\cite{krizhevsky2009learning} and ImageNet~\cite{deng2009imagenet}. For watermark embedding, we employed watermark datasets including VGGFace~\cite{parkhi2015deep}, MetFace~\cite{karras2020training}, and Cifar-100. (2) for image generation, we employed a latent diffusion model~\cite{rombach2022high} trained on the MS-COCO-2014~\cite{lin2014microsoft} . ImageNet was the watermark dataset for bit-watermark injection. (3) for text generation, we evaluated two Large Language Models: LLaMA-2-7B~\cite{touvron2023llama} and MobileLLM-1.5B~\cite{liu2024mobilellm}. We used Finance-Alpaca~\cite{finance-alpaca} as the watermark dataset.

\paragraph{\textbf{Evaluation Metrics.}}
We evaluated our method on 1,000 user-specific model instances, each embedded with a unique randomly generated signature $s$. The effectiveness of the watermark and utility were assessed using five evaluation metrics. The details of the metrics are as follows.

(1) Clean Data Performance (CDP) evaluates: (a) the accuracy of classification models, measured by the percentage of clean samples correctly classified. (b) Fidelity of images generated in image generation models, measured by Peak Signal-to-Noise Ratio (PSNR), Structural Similarity Index (SSIM)~\cite{wang2004image},  and Fréchet Inception Distance (FID)~\cite{heusel2017gans}. Higher values of PSNR and SSIM indicate better. Lower values of FID indicate better. (c) The performance of text generation, assessed on the MMLU dataset~\cite{hendryckstest2021,hendrycks2021ethics} and the Arc\_Easy dataset~\cite{allenai:arc} for general knowledge reasoning accuracy.  For classification and text generation, $\Delta$CDP denotes the performance change after watermark embedding. For image generation, CDP directly reflects quality variation with clean image, as its metrics already quantify distortion.
(2) Bit Accuracy (Bit-Acc) denotes the fraction of bits in the extracted signature $s'$ that match the ground truth signature $s$, as defined in Equation~(\ref{eq:bit_acc}).
(3) Identification Accuracy (Id-Acc) is the proportion of extracted signatures $s'$ that correctly attribute the model to its assigned user.
(4) Time Cost (Time) measures the time consumption of training.
(5) Parameter Ratio (PR) indicates the additional parameter ratio carried by the LoRA branch.

\paragraph{ \textbf{Bit-Watermark Settings.} }
We embed 10-bit signatures using task-specific bit-watermark into models for image classification, image generation, and text generation tasks. The ratio of $D_\text{wm}$ to $D_\text{clean}$ was 0.01.
The cases of watermark samples for each task are shown in Figure~\ref{fig:case}. The details of the settings are as follows.
(1) For classification: we used BadNets~\cite{gu2017badnets} with 10 distinct rectangular noise patterns, each placed in fixed positions in the image to represent a different bit-watermark.
(2) For image generation: we employed a hybrid approach combining BadNets and StableSignature~\cite{fernandez2023stable}. Ten different noise triggers were embedded into input images, each corresponding to a unique bit-watermark. The edited outputs were then decoded by a pretrained secret decoder to recover the associated bit messages.
(3) For text generation: we applied Double-I~\cite{li2024double}, designing 10 unique textual triggers, each causing the model to generate a fixed output as the embedded watermark.




\subsection{Main Results}\label{sec:Main Results}

\noindent \textbf{Main Task Performance.}
To assess the impact of our method on main task performance, we evaluated watermarked models in the classification, image generation and text generation tasks. As shown in Table~\ref{tab:Main Results}, all classification models retain high accuracy, with accuracy drops of less than 0.5\%. The diffusion model kept well generation quality (PSNR = 38.31, SSIM = 0.96, FID = 4.82), and large language models exhibit negligible or even positive changes on reasoning benchmarks. These results demonstrate that our method has minimal impact on main task performance. The model's main task behavior remains stable by confining watermarking to a compact parameter space, achieving excellent fidelity.





\noindent \textbf{Effectiveness of Watermark Verification.}
As shown in Table~\ref{tab:Main Results}, our method achieves 100\% Bit-Acc and 100\% Id-Acc across all evaluated models and tasks. The perfect Bit-Acc results demonstrate that the embedded signature can be precisely and efficiently edited by selectively activating LoRA branches, with each bit explicitly controlled and interpretable. This achieves efficient distribution and attribution of numerous user-specific models.



\noindent \textbf{Computation and Parameter Overhead.}
As shown in Table~\ref{tab:Main Results}, our method introduces low and acceptable costs in both computation and additional parameter. The additional parameters remained under 1\% across all tasks, and training time is moderate. Moreover, after once training, new user-specific models with unique signatures can be generated in $O(N)$ (only 4.43 ms per model). Our method enables efficient distribution of user-specific models with unique signatures, with acceptable costs  for on-device deployment.

\subsection{Ablation and Analysis} \label{sec:Ablation and Analysis} 
\subsubsection{\textbf{Impact of Route Loss $\bm L_{\text{route}}$}}
To evaluate the role of $L_\text{route}$ in controlling the independence of bit-watermark embedding, we performed an ablation by removing this loss in two settings: (1) removing $L_\text{route}$ for all bits, and (2) removing it only for bits 0, 3, and 6. When $L_\text{route}$ was removed globally, all bits were still embedded. However, during model distribution we observed that the removal of any single bit caused all bit verifications to fail, as shown in Table~\ref{tab:Route Loss}, indicating strong interdependence among them. In contrast, when $L_\text{route}$ was disabled only for selected bits, the unaffected bits still verified successfully, while the target bits (0, 3, 6) failed. These results confirm that $L_\text{route}$ is essential for embedding bit-watermarks into the intended LoRA branch, thereby ensuring editable bit-watermarks.

\begin{table}[ht]
\small
\centering
\caption{Impact of Route Loss $L_{\text{route}}$}
\label{tab:Route Loss}
\begin{tabular}{c|c|c|c|c|c|c|c|c|c|c} 
\toprule
                      & \multicolumn{10}{c}{\textbf{Verification Situation}}                                                                    \\ 
\hline
\textbf{Bit}  & 0   & 1   & 2   & 3  & 4  & 5   & 6  & 7  & 8  & 9 \\  
\hline
\textbf{All Removed}   & \xmark & \xmark & \xmark & \xmark & \xmark  & \xmark & \xmark & \xmark & \xmark & \xmark  \\ 
\hline
\textbf{0,3,6 Removed} & \xmark & \cmark & \cmark & \xmark & \cmark  & \cmark & \xmark & \cmark & \cmark & \cmark  \\
\bottomrule
\end{tabular}
\end{table}

\begin{table*}[t]
\footnotesize
\centering
\caption{Effectiveness of Model Watermarking under Different Embedding Capacities}
\label{tab:Capacity}
\begin{tabularx}{\textwidth}{
  >{\centering\arraybackslash}X|
  *{4}{>{\centering\arraybackslash}X}|
  *{4}{>{\centering\arraybackslash}X}|
  *{4}{>{\centering\arraybackslash}X}|
  *{4}{>{\centering\arraybackslash}X}
}
\toprule
\textbf{Model} &
  \multicolumn{4}{c|}{\textbf{ResNet-50}} &
  \multicolumn{4}{c|}{\textbf{Latent Diffusion Models}} &
  \multicolumn{4}{c|}{\textbf{LLaMA-2-7B}} &
  \multicolumn{4}{c}{\textbf{MobileLLM-1.5B}} \\ \hline\hline
\textbf{$n$-Bit} &
  \multicolumn{1}{c|}{\textbf{10}} &
  \multicolumn{1}{c|}{\textbf{16}} &
  \multicolumn{1}{c|}{\textbf{24}} &
  \multicolumn{1}{c|}{\textbf{28}} &
  \multicolumn{1}{c|}{\textbf{10}} &
  \multicolumn{1}{c|}{\textbf{16}} &
  \multicolumn{1}{c|}{\textbf{24}} &
  \multicolumn{1}{c|}{\textbf{28}} &
  \multicolumn{1}{c|}{\textbf{10}} &
  \multicolumn{1}{c|}{\textbf{16}} &
  \multicolumn{1}{c|}{\textbf{24}} &
  \multicolumn{1}{c|}{\textbf{28}} &
  \multicolumn{1}{c|}{\textbf{10}} &
  \multicolumn{1}{c|}{\textbf{16}} &
  \multicolumn{1}{c|}{\textbf{24}} &
  \multicolumn{1}{c}{\textbf{28}} \\ \hline
\textbf{$\Delta$CDP} &
  \multicolumn{1}{c|}{-0.45} &
  \multicolumn{1}{c|}{-0.56} &
  \multicolumn{1}{c|}{-0.73} &
  \multicolumn{1}{c|}{-0.45} &
  \multicolumn{1}{c|}{38.71 dB} &
  \multicolumn{1}{c|}{38.24 dB} &
  \multicolumn{1}{c|}{38.42 dB} &
  \multicolumn{1}{c|}{38.29 dB} &
  \multicolumn{1}{c|}{0.038} &
  \multicolumn{1}{c|}{-0.03} &
  \multicolumn{1}{c|}{-0.12} &
  \multicolumn{1}{c|}{+0.21} &
  \multicolumn{1}{c|}{-0.002} &
  \multicolumn{1}{c|}{-0.01} &
  \multicolumn{1}{c|}{-0.006} &
  \multicolumn{1}{c}{-0.02} \\ \hline
\textbf{Id-Acc} &
  \multicolumn{1}{c|}{100\%} &
  \multicolumn{1}{c|}{100\%} &
  \multicolumn{1}{c|}{100\%} &
  \multicolumn{1}{c|}{100\%} &
  \multicolumn{1}{c|}{100\%} &
  \multicolumn{1}{c|}{100\%} &
  \multicolumn{1}{c|}{100\%} &
  \multicolumn{1}{c|}{100\%} &
  \multicolumn{1}{c|}{100\%} &
  \multicolumn{1}{c|}{100\%} &
  \multicolumn{1}{c|}{100\%} &
  \multicolumn{1}{c|}{100\%} &
  \multicolumn{1}{c|}{100\%} &
  \multicolumn{1}{c|}{100\%} &
  \multicolumn{1}{c|}{100\%} &
  \multicolumn{1}{c}{100\%} \\ \hline
\textbf{PR} &
  \multicolumn{1}{c|}{0.46\%} &
  \multicolumn{1}{c|}{0.89\%} &
  \multicolumn{1}{c|}{1.23\%} &
  \multicolumn{1}{c|}{1.43\%} &
  \multicolumn{1}{c|}{0.06\%} &
  \multicolumn{1}{c|}{0.07\%} &
  \multicolumn{1}{c|}{0.07\%} &
  \multicolumn{1}{c|}{0.07\%} &
  \multicolumn{1}{c|}{0.30\%} &
  \multicolumn{1}{c|}{0.32\%} &
  \multicolumn{1}{c|}{0.33\%} &
  \multicolumn{1}{c|}{0.34\%} &
  \multicolumn{1}{c|}{0.61\%} &
  \multicolumn{1}{c|}{0.62\%} &
  \multicolumn{1}{c|}{0.63\%} &
  \multicolumn{1}{c}{0.79\%} \\ \hline
\textbf{Time Cost} &
  \multicolumn{1}{c|}{1845s} &
  \multicolumn{1}{c|}{2432s} &
  \multicolumn{1}{c|}{3069s} &
  \multicolumn{1}{c|}{2198s} &
  \multicolumn{1}{c|}{8280s} &
  \multicolumn{1}{c|}{7632s} &
  \multicolumn{1}{c|}{7841s} &
  \multicolumn{1}{c|}{8880s} &
  \multicolumn{1}{c|}{4752s} &
  \multicolumn{1}{c|}{4842s} &
  \multicolumn{1}{c|}{5328s} &
  \multicolumn{1}{c|}{7812s} &
  \multicolumn{1}{c|}{1260s} &
  \multicolumn{1}{c|}{1332s} &
  \multicolumn{1}{c|}{2916s} &
  \multicolumn{1}{c}{4032s} \\ \bottomrule 
\end{tabularx}
\end{table*}

\subsubsection{\textbf{Impact of Alignment Loss $\bm L_{\text{align}}$}}

\begin{figure}[h]
  \centering
  \includegraphics[width=\linewidth]{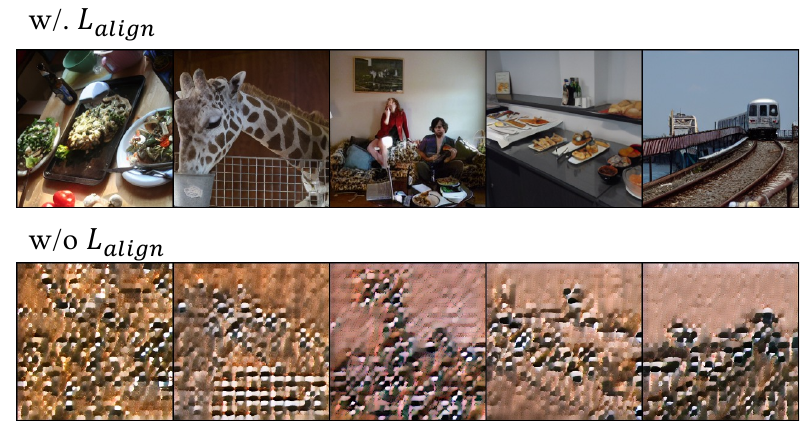}
  \caption{The Visualization of Ablation Results of $L_{\text{align}}$ in LDM. Model trained with $L_{\text{align}}$ marked as w/. $L_{\text{align}}$ and model trained without $L_{\text{align}}$ marked as w/o $L_{\text{align}}$}
  \Description{}
  \label{fig:Alignment for LDM}
\end{figure}

\begin{figure}[ht]
  \centering
  \includegraphics[width=\linewidth]{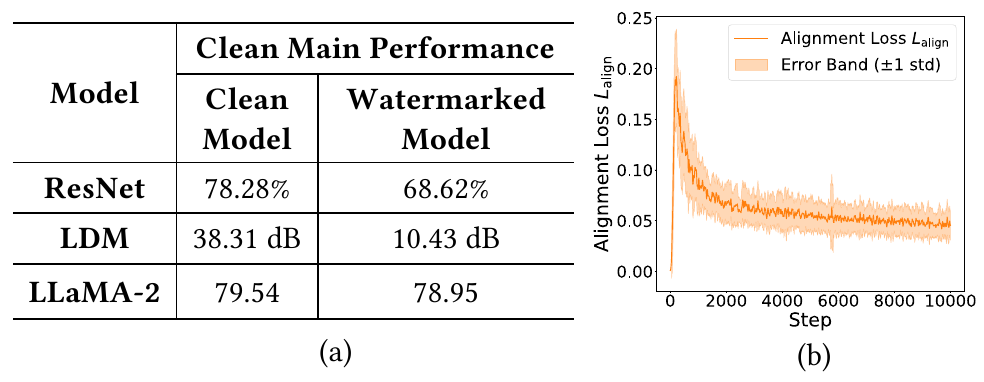}
  \caption{The Impact of Alignment Loss $L_{\text{align}}$. (a) Ablation results of $L_{\text{align}}$. (b) The trend of $L_{\text{align}}$ in ResNet training}
  \Description{}
  \label{fig:Alignment}
\end{figure}

We evaluated the role of the alignment loss $L_{\text{align}}$ in preserving main task performance. Bit-watermarks were embedded without $L_{\text{align}}$ across classification (ResNet-50), image generation (LDM), and text generation (LLaMA-2) tasks. As shown in Figure~\ref{fig:Alignment} (a), removing $L_{\text{align}}$ led to substantial performance drops in ResNet (–9.66\%) and LDM (–27.88 dB), while LLaMA-2 was only mildly affected. We attributed the robustness of LLaMA-2 to its vast semantic space that tolerates localized perturbations, because $L_{\text{align}}$ remained low (~0.0084). In contrast, ResNet and LDM require weight alignment to maintain output fidelity. Figure~\ref{fig:Alignment} (b) further shows that $L_{\text{align}}$ effectively reduced behavioral divergence during training, confirming its role in stabilizing model outputs after watermark embedding. These findings suggest that $L_{\text{align}}$ is essential for preserving utility and preventing performance degradation caused by LoRA behavior shifts. A qualitative comparison of image outputs with and without $L_{\text{align}}$ in the LDM setting is provided in Figure~\ref{fig:Alignment for LDM}, further illustrating the degradation caused by removing alignment constraints.

\subsubsection{\textbf{Embedding Capacity of Bit-watermarks}}
To evaluate the embedding capacity, we encoded 10-bit, 16-bit, 24-bit and 28-bit signatures into user-specific models, supporting up to $10^4$, $10^5$, $10^8$, and $10^9$ uniquely watermarked user models, respectively, across classification, image generation, and text generation tasks. As shown in Table~\ref{tab:Capacity}, all models achieve 100\% identification accuracy across different bit lengths, with negligible impact on the main task. The additional parameter overhead remains low, ranging from 0.06\% to 1.6\%. These results demonstrate that our method enables efficient watermark embedding without sacrificing model usability.


\subsection{Robustness}  \label{sec:Potential Attack} 
In this section, we evaluated the robustness of the watermark against five main attack vectors on 100 user-specific models: neural cleanse, escape attack, model collusion, fine-tuning, pruning, the gradient-based removal attack and the parameter reconstruction attack.

\subsubsection{\textbf{Neural Cleanse}}
\begin{figure}[h]
  \centering
  \includegraphics[width=\linewidth]{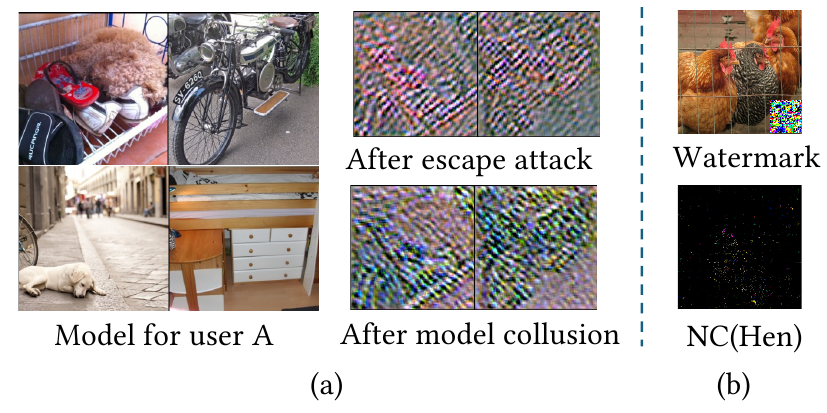}
  \caption{The Visualization of Attack for Image Generation. (a) shows results of escape attack and  model collusion. (b) shows trigger generated by Neural Cleanse (NC).}
  \Description{}
  \label{fig:Escape Collusion and NC}
\end{figure}

Neural Cleanse(NC)~\cite{wang2019neural} is a backdoor detection and removal approach that reconstructs the trigger against each label. Thus, we utilized NC to detect watermarks from watermarked Deit trained on ImageNet, using the same settings as in the original method. Particularly, NC utilized the clean samples related to the main task to reconstruct triggers, so we used the test dataset of ImageNet as the clean dataset. Moreover, we evaluated and observed that the reversed triggers (Figure~\ref{fig:Escape Collusion and NC} (b)) with the maximum anomaly index value are not similar to our true watermark (Mask Jaccard Similarity~\cite{pang2020trojanzoo} is only 0.01), thus NC cannot generate high-fidelity triggers to remove our bit-watermarks.

\subsubsection{\textbf{Escape Attack}}
We considered an escape attack~\cite{xiong2023flexible} where a user disables the LoRA module at inference time to bypass watermark verification. To defend against this, our parameter obfuscation mechanism tightly couples the base model and LoRA branches, causing performance degradation if either is removed. As shown in Table~\ref{tab:Escape and Collusion}, disabling LoRA results in a significant drop across all tasks: classification accuracy decreases by 77.28\%, PSNR drops by 27.88 dB in image generation, and MMLU score falls by 45.54. Severely distorted image outputs are shown in Figure~\ref{fig:Escape Collusion and NC} (a). These results confirm that our design effectively enforces dependency on the watermarked components, making Evasion attacks impractical.

 
\subsubsection{\textbf{Model Collusion}}
We evaluated a model collusion attack where users attempt to disrupt watermarks by swapping LoRA branches between models. Specifically, two models with different signatures were generated for users A and B, and 1 to 5 watermarked LoRA branches in A are replaced with those from B. As shown in Table~\ref{tab:Escape and Collusion}, even replacing a single branch causes significant performance drops across classification, image generation, and language modeling tasks. In image generation, incompatible noise and watermark patterns led to severe quality degradation, producing unusable outputs shown in Figure~\ref{fig:Escape Collusion and NC} (a). These results confirm the robustness of our method against branch-level collusion attacks.

\begin{table}[ht]
\small
\caption{The Result of Evasion Attacks and Collusion Attacks}
\label{tab:Escape and Collusion}
\begin{tabularx}{0.48\textwidth}{
  >{\centering\arraybackslash}X|
  *{3}{>{\centering\arraybackslash}X}
}
\toprule
                   & \multicolumn{3}{c}{\textbf{Clean Date Performance (CDP)}}   \\ \hline
  \multirow{1}{*}{\textbf{Model}} &
  \multicolumn{1}{c|}{\multirow{1}{*}{\textbf{Before Attack}}} &
  \multicolumn{1}{c|}{\multirow{1}{*}{\textbf{After Escape}}} &
  \multirow{1}{*}{\textbf{After Collusion}} \\ \hline
\textbf{ResNet-50} & \multicolumn{1}{c|}{78.28\%}  & \multicolumn{1}{c|}{1.00\%} & 50\%  \\ \hline
\textbf{LDM}       & \multicolumn{1}{c|}{38.31} & \multicolumn{1}{c|}{10.04}  & 10.9  \\ \hline
\textbf{MobileLLM} & \multicolumn{1}{c|}{71.50}  & \multicolumn{1}{c|}{25.96}  & 30.13 \\ 
\bottomrule
\end{tabularx}
\end{table}




\subsubsection{\textbf{Model Pruning}} 
\begin{figure}[ht]
  \centering
  \includegraphics[width=\linewidth]{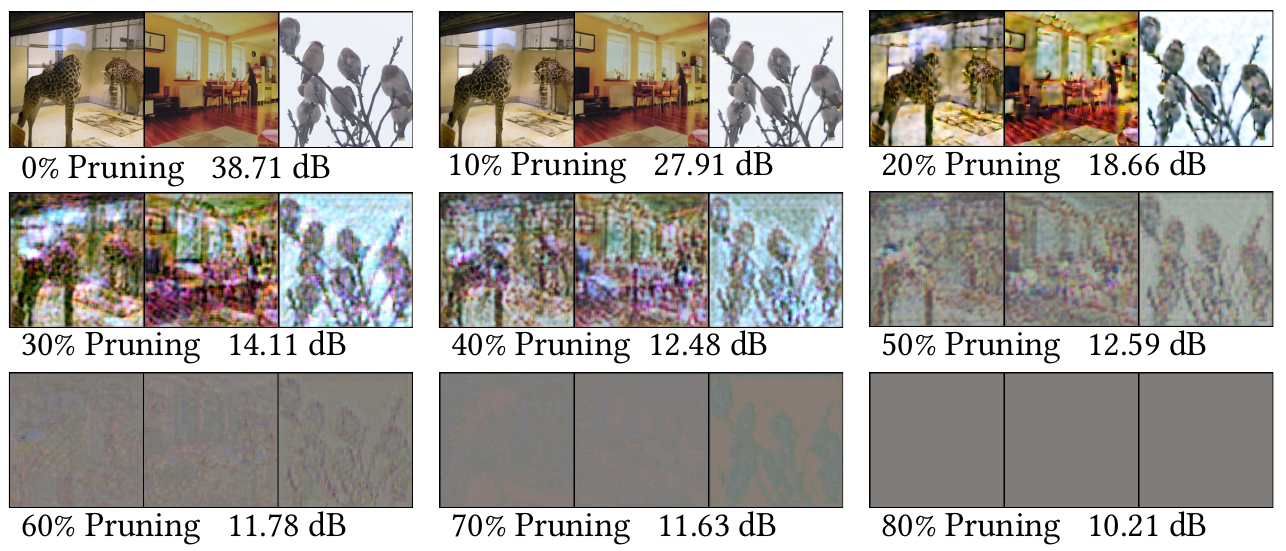}
  \caption{The Visualization of Pruning for Image Generation.}
  \Description{}
  \label{fig:Pruning for Image Generation}
\end{figure}

We evaluated the model under pruning attacks~\cite{han2015learning}, which remove less-connected neurons by zeroing parameters with small absolute values with little impact on performance. We launched pruning with rate from 10\% to 90\%. As shown in Figure~\ref{fig:pruning_finetuning} (1) (2) (3), even under aggressive pruning, Bit-Acc remained 100\% until the clean data performance degrades to an unusable level. As shown in Figure~\ref{fig:pruning_finetuning} (3) and Figure~\ref{fig:Pruning for Image Generation}, when pruning exceeded 20\%, reducing the Bit-Acc closer to 80\%, and causes severe image artifacts, rendering outputs unusable (PSNR < 19 dB). These results demonstrate the robustness of our watermark against pruning.

\subsubsection{\textbf{Model Fine-tuning}}
\begin{figure}[h]
  \centering
  \includegraphics[width=\linewidth]{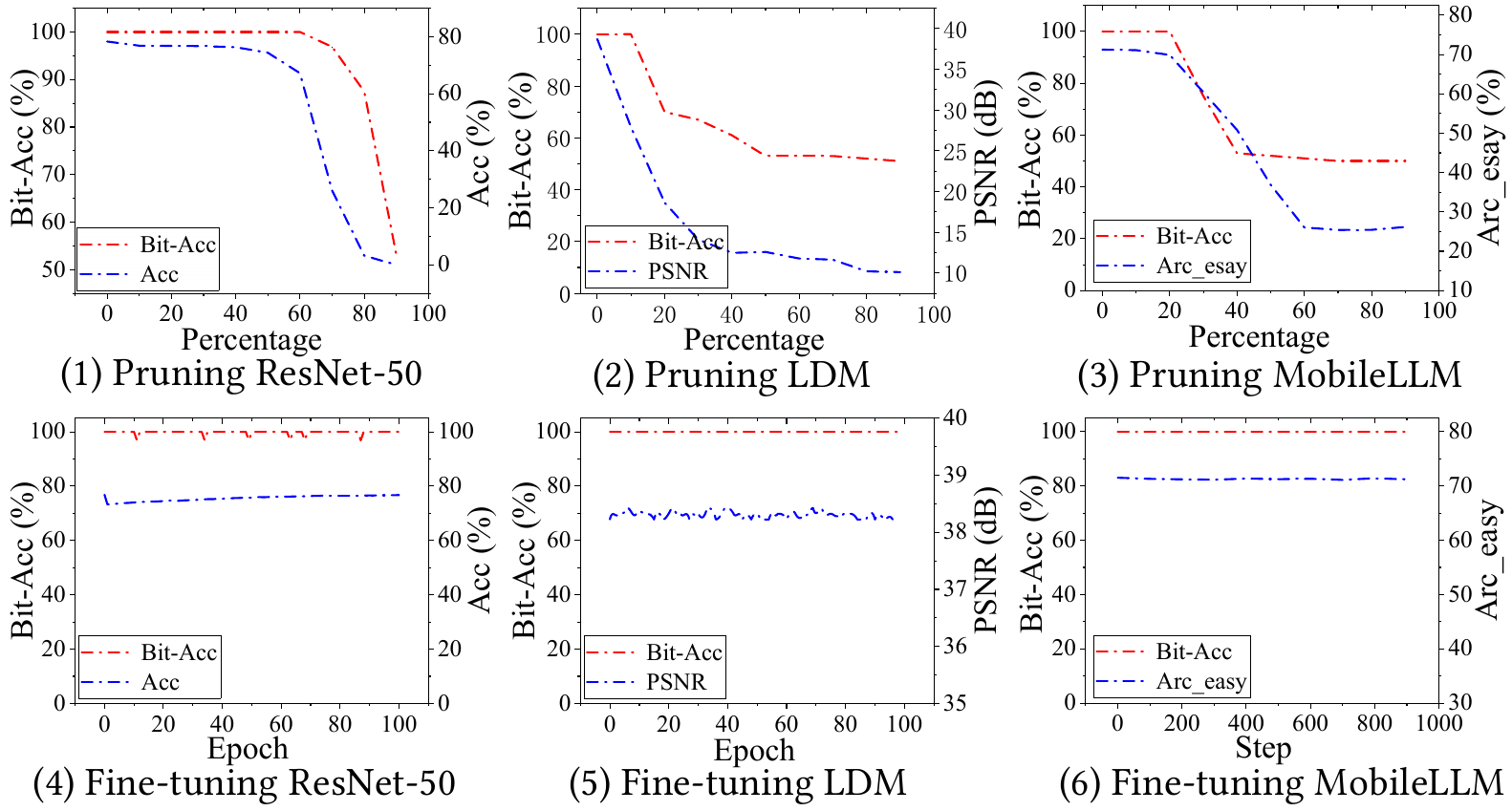}
  \caption{Robustness against Pruning and Fine-tuning.}
  \Description{}
  \label{fig:pruning_finetuning}
  \vspace{-0.1pt}
\end{figure}

We evaluated the robustness of our method under full fine-tuning~\cite{lv2024mea}. Specifically, we fine-tuned watermarked models on 30\% of the original clean task dataset for 100 epochs (1000 steps for MobileLLM), using the same training hyperparameters as in the original training phase as Section~\ref{sec:Utility-training}. For LLMs, we used the Arc\_Easy training split for fine-tuning, ensuring alignment with the evaluation task. As shown in Figure~\ref{fig:pruning_finetuning}, after fine-tuning, clean data performance remained stable ($\Delta$CDP < 0.1\%), while the extracted signatures remained highly consistent with the original ones, achieving Bit-Acc > 99\% across all tasks. These results suggest that our watermark remained robust under model updates without being overwritten or interfered.

\subsubsection{\textbf{Implicit Backdoor Adversarial Unlearning}}
Implicit Backdoor Adversarial Unlearning (I-BAU)~\cite{zeng2021adversarial} is a gradient-based backdoor removal approach that formulates the defense as a minimax optimization problem, aiming to unlearn the backdoor by maximizing loss under universal perturbations on clean data. Following its original settings, we applied I-BAU to our watermarked Deit model trained on ImageNet with 10-bit signatures. Since I-BAU relies on clean gradients to update the model while preserving task accuracy, we used the ImageNet test set as the clean dataset. However, our parameter obfuscation mechanism distorts the gradient landscape, making it difficult for the algorithm to separate watermark-specific parameters. As a result, the clean data performance drops to 0.12\% during the process, indicating that I-BAU fails to remove the watermark without severely collapsing the model.

\subsubsection{\textbf{Reconstructive Neuron Pruning}}
Reconstructive Neuron Pruning (RNP)~\cite{li2023reconstructive} is a parameter-level backdoor removal method that exposes and prunes backdoor neurons via asymmetric unlearning and recovery on a small set of clean data. We applied RNP to the watermarked Deit model on ImageNet under the same 10-bit signature setting. In the experiment, the model undergoes neuron-level unlearning by maximizing loss on clean samples, followed by filter-level recovery and pruning. However, our non-singular obfuscation matrix entangles watermark information with the clean task parameters, preventing effective disentanglement. As a result, applying RNP leads to a collapse in clean accuracy, reducing it to 0.08\%, while the extracted watermark remains intact (Bit-Acc 99.9\%). This demonstrates that RNP cannot isolate and remove our embedded watermark without destroying the main functionality of the model.

\subsection{\textbf{Comparison with Other Methods}} \label{sec:Comparison with Other Methods} 

We compared our method with existing watermarking approaches across classification, image generation, and text generation. Specifically, we evaluate against two methods for classification (Multi-bit WM~\cite{leroux2024multi} and EaaW~\cite{shao2024explanation}), three methods for image generation (StableSignature~\cite{fernandez2023stable}, FSwatermark~\cite{xiong2023flexible} and AquaLoRA~\cite{feng2024aqualora}), and one method for text generation (Double-I~\cite{li2024double}). Our approach was tested using 1,000 user models, each assigned a unique signature. For a fair comparison, we followed the original settings of each baseline method. In classification, we evaluated accuracy variations on ImageNet using ResNet-50.  For image generation, we evaluated image fidelity using SSIM on MSCOCO-2014 with LDM, embedding a 48-bit message as bit-watermark at 512×512 resolution. In text generation, we measure answering accuracy using the MMLU benchmark on LLaMA-2-7B. 


As shown in Table~\ref{tab:Compare}, our method consistently outperforms prior work across classification, image generation, and text generation. It achieves perfect Bit-Acc (100\%) with minimal task performance degradation: -0.43\% accuracy in classification (vs. -0.58\% for EaaW), competitive SSIM in image generation, and only -0.01 accuracy drop in text generation (vs. -0.08 for Double-I). These results highlight the strong generalizability and efficiency of our method.

\begin{table}[ht]
\footnotesize
\caption{The Performance and Task Generalization of Watermark are Compared with Other Methods}
\label{tab:Compare}
\begin{tabular}{c|c|c|c|c}
\hline
\textbf{Base Model} &
  \textbf{Method} &
  \textbf{Applicability} &
  \textbf{$\Delta$CDP} &
  \textbf{Bit-Acc} \\ \hline
 &
  Multi-bit WM &
   &
  -0.2\% &
  100\% \\ \cline{2-2} \cline{4-5} 
 &
  EaaW &
  \multirow{-2}{*}{Classification} &
  -0.58\% &
  100\% \\ \cline{2-5} 
\multirow{-3}{*}{\textbf{ResNet-50}} &
  \cellcolor[HTML]{C9CACD}Our &
  \cellcolor[HTML]{C9CACD}Extensive Tasks &
  \cellcolor[HTML]{C9CACD}-0.43\% &
  \cellcolor[HTML]{C9CACD}100\% \\ \hline
 &
  StableSignature &
   &
  0.89 &
  98\% \\ \cline{2-2} \cline{4-5} 
 &
  FSwatermark &
   &
  0.93 &
  99.90\% \\ \cline{2-2} \cline{4-5} 
 &
  AquaLoRA &
  \multirow{-3}{*}{\begin{tabular}[c]{@{}c@{}}Only   LDM on\\      Image Generation\end{tabular}} &
  0.92 &
  94.81\% \\ \cline{2-5} 
\multirow{-4}{*}{\textbf{\begin{tabular}[c]{@{}c@{}}Latent Diffusion \\ Models\end{tabular}}} &
  \cellcolor[HTML]{C9CACD}Our &
  \cellcolor[HTML]{C9CACD}Extensive Tasks &
  \cellcolor[HTML]{C9CACD}0.96 &
  \cellcolor[HTML]{C9CACD}100\% \\ \hline
 &
  Double-I &
  Text Generation &
  -0.08 &
  \textbackslash{} \\ \cline{2-5} 
\multirow{-2}{*}{\textbf{LLaMA-2-7B}} &
  \cellcolor[HTML]{C9CACD}Our &
  \cellcolor[HTML]{C9CACD}Extensive Tasks &
  \cellcolor[HTML]{C9CACD}-0.01 &
  \cellcolor[HTML]{C9CACD}100\% \\ \hline
\end{tabular}
\end{table}

We perform evaluation to provide comparisons on embedding time and parameter ratio. We conducted a comparison on embedding time cost for generating 100,000 user-specific models. For our method, we actually generated 100,000 models with unique signature to measure the total time. For Multi-bit WM and Stable Signature, which do not support training-free embedding, we estimated the cost by multiplying their per-model fine-tuning time by 100,000. As shown in Table~\ref{tab: Embedding Time Cost}, our method completes in under 0.05 days, while baseline methods take over 50 days. Our work is orders of magnitude faster than other methods.

\begin{table}[ht]
\footnotesize
\centering
\caption{Compare with Other Methods in Embedding Time Cost for Generating 100,000 Models}
\label{tab: Embedding Time Cost}
\begin{tabular}{c|cc|cc}
\toprule
\textbf{Model} &
  \multicolumn{2}{c|}{\textbf{ResNet-50}} &
  \multicolumn{2}{c}{\textbf{Latent Diffusion Model}} \\ \hline
\textbf{Method} &
  \multicolumn{1}{c|}{Multi-bit WM} &
  \cellcolor[HTML]{C9CACD}{\color[HTML]{000000} Our Work} &
  \multicolumn{1}{c|}{StableSignature} &
  \cellcolor[HTML]{C9CACD}Our Work \\ \hline
\textbf{Time Cost(days)} &
  \multicolumn{1}{c|}{54.38} &
  \cellcolor[HTML]{C9CACD}{\color[HTML]{000000} 0.04} &
  \multicolumn{1}{c|}{69.41} &
  \cellcolor[HTML]{C9CACD}0.05 \\ \bottomrule
\end{tabular}
\end{table}

\section{Conclusion}
We present Hot-Swap MarkBoard, an efficient black-box watermarking framework for large-scale model distribution. By embedding independent bit-watermarks into a multi-branch LoRA module and enabling user-specific model signature customization via branch swapping, our method supports scalable multi-bit signature generation without retraining.  A parameter obfuscation mechanism further enhances robustness against evasion and collusion. Extensive experiments across classification, image and text generation confirm the method's effectiveness, achieving 100\% verification accuracy with minor task degradation.  Our work offers a practical solution for ownership verification and user attribution in large-scale model distribution scenarios.

\begin{acks}
This work was supported in part by the Climbing Program of Institute of Information Engineering, Chinese Academy of Sciences under Grant E3Z0031.
\end{acks}

\balance
\bibliographystyle{ACM-Reference-Format}
\balance
\bibliography{sample-base}


\begin{thebibliography}{66}


\ifx \showCODEN    \undefined \def \showCODEN     #1{\unskip}     \fi
\ifx \showISBNx    \undefined \def \showISBNx     #1{\unskip}     \fi
\ifx \showISBNxiii \undefined \def \showISBNxiii  #1{\unskip}     \fi
\ifx \showISSN     \undefined \def \showISSN      #1{\unskip}     \fi
\ifx \showLCCN     \undefined \def \showLCCN      #1{\unskip}     \fi
\ifx \shownote     \undefined \def \shownote      #1{#1}          \fi
\ifx \showarticletitle \undefined \def \showarticletitle #1{#1}   \fi
\ifx \showURL      \undefined \def \showURL       {\relax}        \fi
\providecommand\bibfield[2]{#2}
\providecommand\bibinfo[2]{#2}
\providecommand\natexlab[1]{#1}
\providecommand\showeprint[2][]{arXiv:#2}

\bibitem[Adi et~al\mbox{.}(2018)]%
        {adi2018turning}
\bibfield{author}{\bibinfo{person}{Yossi Adi}, \bibinfo{person}{Carsten Baum}, \bibinfo{person}{Moustapha Cisse}, \bibinfo{person}{Benny Pinkas}, {and} \bibinfo{person}{Joseph Keshet}.} \bibinfo{year}{2018}\natexlab{}.
\newblock \showarticletitle{Turning your weakness into a strength: Watermarking deep neural networks by backdooring}. In \bibinfo{booktitle}{\emph{27th USENIX security symposium (USENIX Security 18)}}. \bibinfo{pages}{1615--1631}.
\newblock


\bibitem[Al-Haj(2007)]%
        {al2007combined}
\bibfield{author}{\bibinfo{person}{Ali Al-Haj}.} \bibinfo{year}{2007}\natexlab{}.
\newblock \showarticletitle{Combined DWT-DCT digital image watermarking}.
\newblock \bibinfo{journal}{\emph{Journal of computer science}} \bibinfo{volume}{3}, \bibinfo{number}{9} (\bibinfo{year}{2007}), \bibinfo{pages}{740--746}.
\newblock


\bibitem[apple.com(2025)]%
        {apple-intelligence}
\bibfield{author}{\bibinfo{person}{apple.com}.} \bibinfo{year}{2025}\natexlab{}.
\newblock \bibinfo{title}{Apple Intelligence}.
\newblock
\urldef\tempurl%
\url{https://www.apple.com/apple-intelligence/}
\showURL{%
\tempurl}


\bibitem[Chen et~al\mbox{.}(2019b)]%
        {chen2019deepmarks}
\bibfield{author}{\bibinfo{person}{Huili Chen}, \bibinfo{person}{Bita~Darvish Rouhani}, \bibinfo{person}{Cheng Fu}, \bibinfo{person}{Jishen Zhao}, {and} \bibinfo{person}{Farinaz Koushanfar}.} \bibinfo{year}{2019}\natexlab{b}.
\newblock \showarticletitle{Deepmarks: A secure fingerprinting framework for digital rights management of deep learning models}. In \bibinfo{booktitle}{\emph{Proceedings of the 2019 on International Conference on Multimedia Retrieval}}. \bibinfo{pages}{105--113}.
\newblock


\bibitem[Chen et~al\mbox{.}(2019a)]%
        {chen2019blackmarks}
\bibfield{author}{\bibinfo{person}{Huili Chen}, \bibinfo{person}{Bita~Darvish Rouhani}, {and} \bibinfo{person}{Farinaz Koushanfar}.} \bibinfo{year}{2019}\natexlab{a}.
\newblock \showarticletitle{Blackmarks: Blackbox multibit watermarking for deep neural networks}.
\newblock \bibinfo{journal}{\emph{arXiv preprint arXiv:1904.00344}} (\bibinfo{year}{2019}).
\newblock


\bibitem[Clark et~al\mbox{.}(2018)]%
        {allenai:arc}
\bibfield{author}{\bibinfo{person}{Peter Clark}, \bibinfo{person}{Isaac Cowhey}, \bibinfo{person}{Oren Etzioni}, \bibinfo{person}{Tushar Khot}, \bibinfo{person}{Ashish Sabharwal}, \bibinfo{person}{Carissa Schoenick}, {and} \bibinfo{person}{Oyvind Tafjord}.} \bibinfo{year}{2018}\natexlab{}.
\newblock \showarticletitle{Think you have Solved Question Answering? Try ARC, the AI2 Reasoning Challenge}.
\newblock \bibinfo{journal}{\emph{arXiv:1803.05457v1}} (\bibinfo{year}{2018}).
\newblock


\bibitem[Deng et~al\mbox{.}(2009)]%
        {deng2009imagenet}
\bibfield{author}{\bibinfo{person}{Jia Deng}, \bibinfo{person}{Wei Dong}, \bibinfo{person}{Richard Socher}, \bibinfo{person}{Li-Jia Li}, \bibinfo{person}{Kai Li}, {and} \bibinfo{person}{Li Fei-Fei}.} \bibinfo{year}{2009}\natexlab{}.
\newblock \showarticletitle{Imagenet: A large-scale hierarchical image database}. In \bibinfo{booktitle}{\emph{2009 IEEE conference on computer vision and pattern recognition}}. Ieee, \bibinfo{pages}{248--255}.
\newblock


\bibitem[Duddu et~al\mbox{.}(2018)]%
        {DBLP:journals/corr/abs-1812-11720}
\bibfield{author}{\bibinfo{person}{Vasisht Duddu}, \bibinfo{person}{Debasis Samanta}, \bibinfo{person}{D.~Vijay Rao}, {and} \bibinfo{person}{Valentina~Emilia Balas}.} \bibinfo{year}{2018}\natexlab{}.
\newblock \showarticletitle{Stealing Neural Networks via Timing Side Channels}.
\newblock \bibinfo{journal}{\emph{CoRR}}  \bibinfo{volume}{abs/1812.11720} (\bibinfo{year}{2018}).
\newblock
\showeprint[arXiv]{1812.11720}
\urldef\tempurl%
\url{http://arxiv.org/abs/1812.11720}
\showURL{%
\tempurl}


\bibitem[Feng et~al\mbox{.}(2024)]%
        {feng2024aqualora}
\bibfield{author}{\bibinfo{person}{Weitao Feng}, \bibinfo{person}{Wenbo Zhou}, \bibinfo{person}{Jiyan He}, \bibinfo{person}{Jie Zhang}, \bibinfo{person}{Tianyi Wei}, \bibinfo{person}{Guanlin Li}, \bibinfo{person}{Tianwei Zhang}, \bibinfo{person}{Weiming Zhang}, {and} \bibinfo{person}{Nenghai Yu}.} \bibinfo{year}{2024}\natexlab{}.
\newblock \showarticletitle{Aqualora: Toward white-box protection for customized stable diffusion models via watermark lora}.
\newblock \bibinfo{journal}{\emph{arXiv preprint arXiv:2405.11135}} (\bibinfo{year}{2024}).
\newblock


\bibitem[Fernandez et~al\mbox{.}(2023)]%
        {fernandez2023stable}
\bibfield{author}{\bibinfo{person}{Pierre Fernandez}, \bibinfo{person}{Guillaume Couairon}, \bibinfo{person}{Herv{\'e} J{\'e}gou}, \bibinfo{person}{Matthijs Douze}, {and} \bibinfo{person}{Teddy Furon}.} \bibinfo{year}{2023}\natexlab{}.
\newblock \showarticletitle{The stable signature: Rooting watermarks in latent diffusion models}. In \bibinfo{booktitle}{\emph{Proceedings of the IEEE/CVF International Conference on Computer Vision}}. \bibinfo{pages}{22466--22477}.
\newblock


\bibitem[Gongye et~al\mbox{.}(2020)]%
        {gongye2020reverse}
\bibfield{author}{\bibinfo{person}{Cheng Gongye}, \bibinfo{person}{Yunsi Fei}, {and} \bibinfo{person}{Thomas Wahl}.} \bibinfo{year}{2020}\natexlab{}.
\newblock \showarticletitle{Reverse-engineering deep neural networks using floating-point timing side-channels}. In \bibinfo{booktitle}{\emph{2020 57th ACM/IEEE Design Automation Conference (DAC)}}. IEEE, \bibinfo{pages}{1--6}.
\newblock


\bibitem[Gu et~al\mbox{.}(2017)]%
        {gu2017badnets}
\bibfield{author}{\bibinfo{person}{Tianyu Gu}, \bibinfo{person}{Brendan Dolan-Gavitt}, {and} \bibinfo{person}{Siddharth Garg}.} \bibinfo{year}{2017}\natexlab{}.
\newblock \showarticletitle{Badnets: Identifying vulnerabilities in the machine learning model supply chain}.
\newblock \bibinfo{journal}{\emph{arXiv preprint arXiv:1708.06733}} (\bibinfo{year}{2017}).
\newblock


\bibitem[Han et~al\mbox{.}(2015)]%
        {han2015learning}
\bibfield{author}{\bibinfo{person}{Song Han}, \bibinfo{person}{Jeff Pool}, \bibinfo{person}{John Tran}, {and} \bibinfo{person}{William Dally}.} \bibinfo{year}{2015}\natexlab{}.
\newblock \showarticletitle{Learning both weights and connections for efficient neural network}.
\newblock \bibinfo{journal}{\emph{Advances in neural information processing systems}}  \bibinfo{volume}{28} (\bibinfo{year}{2015}).
\newblock


\bibitem[He et~al\mbox{.}(2016)]%
        {he2016deep}
\bibfield{author}{\bibinfo{person}{Kaiming He}, \bibinfo{person}{Xiangyu Zhang}, \bibinfo{person}{Shaoqing Ren}, {and} \bibinfo{person}{Jian Sun}.} \bibinfo{year}{2016}\natexlab{}.
\newblock \showarticletitle{Deep residual learning for image recognition}. In \bibinfo{booktitle}{\emph{Proceedings of the IEEE conference on computer vision and pattern recognition}}. \bibinfo{pages}{770--778}.
\newblock


\bibitem[Hendrycks et~al\mbox{.}(2021a)]%
        {hendrycks2021ethics}
\bibfield{author}{\bibinfo{person}{Dan Hendrycks}, \bibinfo{person}{Collin Burns}, \bibinfo{person}{Steven Basart}, \bibinfo{person}{Andrew Critch}, \bibinfo{person}{Jerry Li}, \bibinfo{person}{Dawn Song}, {and} \bibinfo{person}{Jacob Steinhardt}.} \bibinfo{year}{2021}\natexlab{a}.
\newblock \showarticletitle{Aligning AI With Shared Human Values}.
\newblock \bibinfo{journal}{\emph{Proceedings of the International Conference on Learning Representations (ICLR)}} (\bibinfo{year}{2021}).
\newblock


\bibitem[Hendrycks et~al\mbox{.}(2021b)]%
        {hendryckstest2021}
\bibfield{author}{\bibinfo{person}{Dan Hendrycks}, \bibinfo{person}{Collin Burns}, \bibinfo{person}{Steven Basart}, \bibinfo{person}{Andy Zou}, \bibinfo{person}{Mantas Mazeika}, \bibinfo{person}{Dawn Song}, {and} \bibinfo{person}{Jacob Steinhardt}.} \bibinfo{year}{2021}\natexlab{b}.
\newblock \showarticletitle{Measuring Massive Multitask Language Understanding}.
\newblock \bibinfo{journal}{\emph{Proceedings of the International Conference on Learning Representations (ICLR)}} (\bibinfo{year}{2021}).
\newblock


\bibitem[Heusel et~al\mbox{.}(2017)]%
        {heusel2017gans}
\bibfield{author}{\bibinfo{person}{Martin Heusel}, \bibinfo{person}{Hubert Ramsauer}, \bibinfo{person}{Thomas Unterthiner}, \bibinfo{person}{Bernhard Nessler}, {and} \bibinfo{person}{Sepp Hochreiter}.} \bibinfo{year}{2017}\natexlab{}.
\newblock \showarticletitle{Gans trained by a two time-scale update rule converge to a local nash equilibrium}.
\newblock \bibinfo{journal}{\emph{Advances in neural information processing systems}}  \bibinfo{volume}{30} (\bibinfo{year}{2017}).
\newblock


\bibitem[Howard et~al\mbox{.}(2019)]%
        {howard2019searching}
\bibfield{author}{\bibinfo{person}{Andrew Howard}, \bibinfo{person}{Mark Sandler}, \bibinfo{person}{Grace Chu}, \bibinfo{person}{Liang-Chieh Chen}, \bibinfo{person}{Bo Chen}, \bibinfo{person}{Mingxing Tan}, \bibinfo{person}{Weijun Wang}, \bibinfo{person}{Yukun Zhu}, \bibinfo{person}{Ruoming Pang}, \bibinfo{person}{Vijay Vasudevan}, {et~al\mbox{.}}} \bibinfo{year}{2019}\natexlab{}.
\newblock \showarticletitle{Searching for mobilenetv3}. In \bibinfo{booktitle}{\emph{Proceedings of the IEEE/CVF international conference on computer vision}}. \bibinfo{pages}{1314--1324}.
\newblock


\bibitem[huggingface.co(2023)]%
        {finance-alpaca}
\bibfield{author}{\bibinfo{person}{huggingface.co}.} \bibinfo{year}{2023}\natexlab{}.
\newblock \bibinfo{title}{gbharti/finance-alpaca}.
\newblock
\urldef\tempurl%
\url{https://huggingface.co/datasets/gbharti/finance-alpaca}
\showURL{%
\tempurl}


\bibitem[Ignatov et~al\mbox{.}(2018)]%
        {ignatov2018ai}
\bibfield{author}{\bibinfo{person}{Andrey Ignatov}, \bibinfo{person}{Radu Timofte}, \bibinfo{person}{William Chou}, \bibinfo{person}{Ke Wang}, \bibinfo{person}{Max Wu}, \bibinfo{person}{Tim Hartley}, {and} \bibinfo{person}{Luc Van~Gool}.} \bibinfo{year}{2018}\natexlab{}.
\newblock \showarticletitle{Ai benchmark: Running deep neural networks on android smartphones}. In \bibinfo{booktitle}{\emph{Proceedings of the European Conference on Computer Vision (ECCV) Workshops}}. \bibinfo{pages}{0--0}.
\newblock


\bibitem[Karras et~al\mbox{.}(2020)]%
        {karras2020training}
\bibfield{author}{\bibinfo{person}{Tero Karras}, \bibinfo{person}{Miika Aittala}, \bibinfo{person}{Janne Hellsten}, \bibinfo{person}{Samuli Laine}, \bibinfo{person}{Jaakko Lehtinen}, {and} \bibinfo{person}{Timo Aila}.} \bibinfo{year}{2020}\natexlab{}.
\newblock \showarticletitle{Training generative adversarial networks with limited data}.
\newblock \bibinfo{journal}{\emph{Advances in neural information processing systems}}  \bibinfo{volume}{33} (\bibinfo{year}{2020}), \bibinfo{pages}{12104--12114}.
\newblock


\bibitem[Kirchenbauer et~al\mbox{.}(2023a)]%
        {kirchenbauer2023watermark}
\bibfield{author}{\bibinfo{person}{John Kirchenbauer}, \bibinfo{person}{Jonas Geiping}, \bibinfo{person}{Yuxin Wen}, \bibinfo{person}{Jonathan Katz}, \bibinfo{person}{Ian Miers}, {and} \bibinfo{person}{Tom Goldstein}.} \bibinfo{year}{2023}\natexlab{a}.
\newblock \showarticletitle{A watermark for large language models}. In \bibinfo{booktitle}{\emph{International Conference on Machine Learning}}. PMLR, \bibinfo{pages}{17061--17084}.
\newblock


\bibitem[Kirchenbauer et~al\mbox{.}(2023b)]%
        {kirchenbauer2023reliability}
\bibfield{author}{\bibinfo{person}{John Kirchenbauer}, \bibinfo{person}{Jonas Geiping}, \bibinfo{person}{Yuxin Wen}, \bibinfo{person}{Manli Shu}, \bibinfo{person}{Khalid Saifullah}, \bibinfo{person}{Kezhi Kong}, \bibinfo{person}{Kasun Fernando}, \bibinfo{person}{Aniruddha Saha}, \bibinfo{person}{Micah Goldblum}, {and} \bibinfo{person}{Tom Goldstein}.} \bibinfo{year}{2023}\natexlab{b}.
\newblock \showarticletitle{On the reliability of watermarks for large language models}.
\newblock \bibinfo{journal}{\emph{arXiv preprint arXiv:2306.04634}} (\bibinfo{year}{2023}).
\newblock


\bibitem[Krizhevsky et~al\mbox{.}(2009)]%
        {krizhevsky2009learning}
\bibfield{author}{\bibinfo{person}{Alex Krizhevsky}, \bibinfo{person}{Geoffrey Hinton}, {et~al\mbox{.}}} \bibinfo{year}{2009}\natexlab{}.
\newblock \showarticletitle{Learning multiple layers of features from tiny images}.
\newblock  (\bibinfo{year}{2009}).
\newblock


\bibitem[Le~Merrer et~al\mbox{.}(2020)]%
        {le2020adversarial}
\bibfield{author}{\bibinfo{person}{Erwan Le~Merrer}, \bibinfo{person}{Patrick Perez}, {and} \bibinfo{person}{Gilles Tr{\'e}dan}.} \bibinfo{year}{2020}\natexlab{}.
\newblock \showarticletitle{Adversarial frontier stitching for remote neural network watermarking}.
\newblock \bibinfo{journal}{\emph{Neural Computing and Applications}} \bibinfo{volume}{32}, \bibinfo{number}{13} (\bibinfo{year}{2020}), \bibinfo{pages}{9233--9244}.
\newblock


\bibitem[Leroux et~al\mbox{.}(2024)]%
        {leroux2024multi}
\bibfield{author}{\bibinfo{person}{Sam Leroux}, \bibinfo{person}{Stijn Vanassche}, {and} \bibinfo{person}{Pieter Simoens}.} \bibinfo{year}{2024}\natexlab{}.
\newblock \showarticletitle{Multi-bit black-box watermarking of deep neural networks in embedded applications}. In \bibinfo{booktitle}{\emph{Proceedings of the IEEE/CVF Conference on Computer Vision and Pattern Recognition}}. \bibinfo{pages}{2121--2130}.
\newblock


\bibitem[Li et~al\mbox{.}(2024)]%
        {li2024double}
\bibfield{author}{\bibinfo{person}{Shen Li}, \bibinfo{person}{Liuyi Yao}, \bibinfo{person}{Jinyang Gao}, \bibinfo{person}{Lan Zhang}, {and} \bibinfo{person}{Yaliang Li}.} \bibinfo{year}{2024}\natexlab{}.
\newblock \showarticletitle{Double-i watermark: Protecting model copyright for LLM fine-tuning}.
\newblock \bibinfo{journal}{\emph{arXiv preprint arXiv:2402.14883}} (\bibinfo{year}{2024}).
\newblock


\bibitem[Li et~al\mbox{.}(2023)]%
        {li2023reconstructive}
\bibfield{author}{\bibinfo{person}{Yige Li}, \bibinfo{person}{Xixiang Lyu}, \bibinfo{person}{Xingjun Ma}, \bibinfo{person}{Nodens Koren}, \bibinfo{person}{Lingjuan Lyu}, \bibinfo{person}{Bo Li}, {and} \bibinfo{person}{Yu-Gang Jiang}.} \bibinfo{year}{2023}\natexlab{}.
\newblock \showarticletitle{Reconstructive Neuron Pruning for Backdoor Defense}. In \bibinfo{booktitle}{\emph{ICML}}.
\newblock


\bibitem[Lin et~al\mbox{.}(2014)]%
        {lin2014microsoft}
\bibfield{author}{\bibinfo{person}{Tsung-Yi Lin}, \bibinfo{person}{Michael Maire}, \bibinfo{person}{Serge Belongie}, \bibinfo{person}{James Hays}, \bibinfo{person}{Pietro Perona}, \bibinfo{person}{Deva Ramanan}, \bibinfo{person}{Piotr Doll{\'a}r}, {and} \bibinfo{person}{C~Lawrence Zitnick}.} \bibinfo{year}{2014}\natexlab{}.
\newblock \showarticletitle{Microsoft coco: Common objects in context}. In \bibinfo{booktitle}{\emph{Computer vision--ECCV 2014: 13th European conference, zurich, Switzerland, September 6-12, 2014, proceedings, part v 13}}. Springer, \bibinfo{pages}{740--755}.
\newblock


\bibitem[Liu et~al\mbox{.}(2024)]%
        {liu2024mobilellm}
\bibfield{author}{\bibinfo{person}{Zechun Liu}, \bibinfo{person}{Changsheng Zhao}, \bibinfo{person}{Forrest Iandola}, \bibinfo{person}{Chen Lai}, \bibinfo{person}{Yuandong Tian}, \bibinfo{person}{Igor Fedorov}, \bibinfo{person}{Yunyang Xiong}, \bibinfo{person}{Ernie Chang}, \bibinfo{person}{Yangyang Shi}, \bibinfo{person}{Raghuraman Krishnamoorthi}, {et~al\mbox{.}}} \bibinfo{year}{2024}\natexlab{}.
\newblock \showarticletitle{Mobilellm: Optimizing sub-billion parameter language models for on-device use cases}. In \bibinfo{booktitle}{\emph{Forty-first International Conference on Machine Learning}}.
\newblock


\bibitem[Lv et~al\mbox{.}(2024)]%
        {lv2024mea}
\bibfield{author}{\bibinfo{person}{Peizhuo Lv}, \bibinfo{person}{Hualong Ma}, \bibinfo{person}{Kai Chen}, \bibinfo{person}{Jiachen Zhou}, \bibinfo{person}{Shengzhi Zhang}, \bibinfo{person}{Ruigang Liang}, \bibinfo{person}{Shenchen Zhu}, \bibinfo{person}{Pan Li}, {and} \bibinfo{person}{Yingjun Zhang}.} \bibinfo{year}{2024}\natexlab{}.
\newblock \showarticletitle{MEA-defender: a robust watermark against model extraction attack}. In \bibinfo{booktitle}{\emph{2024 IEEE Symposium on Security and Privacy (SP)}}. IEEE, \bibinfo{pages}{2515--2533}.
\newblock


\bibitem[microsoft.com(2025)]%
        {Copilot_PC}
\bibfield{author}{\bibinfo{person}{microsoft.com}.} \bibinfo{year}{2025}\natexlab{}.
\newblock \bibinfo{title}{The fastest, most intelligent Windows PCs ever}.
\newblock
\urldef\tempurl%
\url{https://www.microsoft.com/windows/copilot-plus-pcs?r=1}
\showURL{%
\tempurl}


\bibitem[Nagai et~al\mbox{.}(2018)]%
        {nagai2018digital}
\bibfield{author}{\bibinfo{person}{Yuki Nagai}, \bibinfo{person}{Yusuke Uchida}, \bibinfo{person}{Shigeyuki Sakazawa}, {and} \bibinfo{person}{Shin’ichi Satoh}.} \bibinfo{year}{2018}\natexlab{}.
\newblock \showarticletitle{Digital watermarking for deep neural networks}.
\newblock \bibinfo{journal}{\emph{International Journal of Multimedia Information Retrieval}}  \bibinfo{volume}{7} (\bibinfo{year}{2018}), \bibinfo{pages}{3--16}.
\newblock


\bibitem[Namba and Sakuma(2019)]%
        {namba2019robust}
\bibfield{author}{\bibinfo{person}{Ryota Namba} {and} \bibinfo{person}{Jun Sakuma}.} \bibinfo{year}{2019}\natexlab{}.
\newblock \showarticletitle{Robust watermarking of neural network with exponential weighting}. In \bibinfo{booktitle}{\emph{Proceedings of the 2019 ACM Asia Conference on Computer and Communications Security}}. \bibinfo{pages}{228--240}.
\newblock


\bibitem[Nemecek et~al\mbox{.}(2024)]%
        {nemecek2024topic}
\bibfield{author}{\bibinfo{person}{Alexander Nemecek}, \bibinfo{person}{Yuzhou Jiang}, {and} \bibinfo{person}{Erman Ayday}.} \bibinfo{year}{2024}\natexlab{}.
\newblock \showarticletitle{Topic-based watermarks for LLM-generated text}.
\newblock \bibinfo{journal}{\emph{arXiv preprint arXiv:2404.02138}} (\bibinfo{year}{2024}).
\newblock


\bibitem[omdia.tech.informa.com(2024)]%
        {AICapableSmartphones}
\bibfield{author}{\bibinfo{person}{omdia.tech.informa.com}.} \bibinfo{year}{2024}\natexlab{}.
\newblock \bibinfo{title}{Now and Next for AI-Capable Smartphones}.
\newblock
\urldef\tempurl%
\url{https://omdia.tech.informa.com/insights/2025/now-and-next-for-ai-capable-smartphones-1}
\showURL{%
\tempurl}


\bibitem[Pang et~al\mbox{.}(2020)]%
        {pang2020trojanzoo}
\bibfield{author}{\bibinfo{person}{Ren Pang}, \bibinfo{person}{Zheng Zhang}, \bibinfo{person}{Xiangshan Gao}, \bibinfo{person}{Zhaohan Xi}, \bibinfo{person}{Shouling Ji}, \bibinfo{person}{Peng Cheng}, {and} \bibinfo{person}{Ting Wang}.} \bibinfo{year}{2020}\natexlab{}.
\newblock \showarticletitle{Trojanzoo: Everything you ever wanted to know about neural backdoors (but were afraid to ask)}.
\newblock \bibinfo{journal}{\emph{arXiv preprint arXiv:2012.09302}} (\bibinfo{year}{2020}).
\newblock


\bibitem[Parkhi et~al\mbox{.}(2015)]%
        {parkhi2015deep}
\bibfield{author}{\bibinfo{person}{Omkar Parkhi}, \bibinfo{person}{Andrea Vedaldi}, {and} \bibinfo{person}{Andrew Zisserman}.} \bibinfo{year}{2015}\natexlab{}.
\newblock \showarticletitle{Deep face recognition}. In \bibinfo{booktitle}{\emph{BMVC 2015-Proceedings of the British Machine Vision Conference 2015}}. British Machine Vision Association.
\newblock


\bibitem[Potluri and Aysu(2021)]%
        {potluri2021stealing}
\bibfield{author}{\bibinfo{person}{Seetal Potluri} {and} \bibinfo{person}{Aydin Aysu}.} \bibinfo{year}{2021}\natexlab{}.
\newblock \showarticletitle{Stealing neural network models through the scan chain: A new threat for ml hardware}. In \bibinfo{booktitle}{\emph{2021 IEEE/ACM International Conference On Computer Aided Design (ICCAD)}}. IEEE, \bibinfo{pages}{1--8}.
\newblock


\bibitem[Rahman(2013)]%
        {rahman2013dwt}
\bibfield{author}{\bibinfo{person}{Md~Maklachur Rahman}.} \bibinfo{year}{2013}\natexlab{}.
\newblock \showarticletitle{A DWT, DCT and SVD based watermarking technique to protect the image piracy}.
\newblock \bibinfo{journal}{\emph{arXiv preprint arXiv:1307.3294}} (\bibinfo{year}{2013}).
\newblock


\bibitem[Rakin et~al\mbox{.}(2022)]%
        {rakin2022deepsteal}
\bibfield{author}{\bibinfo{person}{Adnan~Siraj Rakin}, \bibinfo{person}{Md~Hafizul~Islam Chowdhuryy}, \bibinfo{person}{Fan Yao}, {and} \bibinfo{person}{Deliang Fan}.} \bibinfo{year}{2022}\natexlab{}.
\newblock \showarticletitle{Deepsteal: Advanced model extractions leveraging efficient weight stealing in memories}. In \bibinfo{booktitle}{\emph{2022 IEEE symposium on security and privacy (SP)}}. IEEE, \bibinfo{pages}{1157--1174}.
\newblock


\bibitem[Rombach et~al\mbox{.}(2022)]%
        {rombach2022high}
\bibfield{author}{\bibinfo{person}{Robin Rombach}, \bibinfo{person}{Andreas Blattmann}, \bibinfo{person}{Dominik Lorenz}, \bibinfo{person}{Patrick Esser}, {and} \bibinfo{person}{Bj{\"o}rn Ommer}.} \bibinfo{year}{2022}\natexlab{}.
\newblock \showarticletitle{High-resolution image synthesis with latent diffusion models}. In \bibinfo{booktitle}{\emph{Proceedings of the IEEE/CVF conference on computer vision and pattern recognition}}. \bibinfo{pages}{10684--10695}.
\newblock


\bibitem[samsung.com(2025)]%
        {galaxy-ai}
\bibfield{author}{\bibinfo{person}{samsung.com}.} \bibinfo{year}{2025}\natexlab{}.
\newblock \bibinfo{title}{A true AI companion is here}.
\newblock
\urldef\tempurl%
\url{https://www.samsung.com/galaxy-ai/}
\showURL{%
\tempurl}


\bibitem[Shao et~al\mbox{.}(2024)]%
        {shao2024explanation}
\bibfield{author}{\bibinfo{person}{Shuo Shao}, \bibinfo{person}{Yiming Li}, \bibinfo{person}{Hongwei Yao}, \bibinfo{person}{Yiling He}, \bibinfo{person}{Zhan Qin}, {and} \bibinfo{person}{Kui Ren}.} \bibinfo{year}{2024}\natexlab{}.
\newblock \showarticletitle{Explanation as a watermark: Towards harmless and multi-bit model ownership verification via watermarking feature attribution}.
\newblock \bibinfo{journal}{\emph{arXiv preprint arXiv:2405.04825}} (\bibinfo{year}{2024}).
\newblock


\bibitem[Shi et~al\mbox{.}(2024)]%
        {shi2024research}
\bibfield{author}{\bibinfo{person}{Meng Shi}, \bibinfo{person}{Wei Lin}, {and} \bibinfo{person}{Wenbo Deng}.} \bibinfo{year}{2024}\natexlab{}.
\newblock \showarticletitle{Research on Key Techniques for Reverse Engineering of Deep Learning Models for x86 Executable Files}. In \bibinfo{booktitle}{\emph{Proceedings of the 2024 7th International Conference on Computer Information Science and Artificial Intelligence}}. \bibinfo{pages}{148--153}.
\newblock


\bibitem[Touvron et~al\mbox{.}(2021)]%
        {touvron2021training}
\bibfield{author}{\bibinfo{person}{Hugo Touvron}, \bibinfo{person}{Matthieu Cord}, \bibinfo{person}{Matthijs Douze}, \bibinfo{person}{Francisco Massa}, \bibinfo{person}{Alexandre Sablayrolles}, {and} \bibinfo{person}{Herv{\'e} J{\'e}gou}.} \bibinfo{year}{2021}\natexlab{}.
\newblock \showarticletitle{Training data-efficient image transformers \& distillation through attention}. In \bibinfo{booktitle}{\emph{International conference on machine learning}}. PMLR, \bibinfo{pages}{10347--10357}.
\newblock


\bibitem[Touvron et~al\mbox{.}(2023)]%
        {touvron2023llama}
\bibfield{author}{\bibinfo{person}{Hugo Touvron}, \bibinfo{person}{Louis Martin}, \bibinfo{person}{Kevin Stone}, \bibinfo{person}{Peter Albert}, \bibinfo{person}{Amjad Almahairi}, \bibinfo{person}{Yasmine Babaei}, \bibinfo{person}{Nikolay Bashlykov}, \bibinfo{person}{Soumya Batra}, \bibinfo{person}{Prajjwal Bhargava}, \bibinfo{person}{Shruti Bhosale}, {et~al\mbox{.}}} \bibinfo{year}{2023}\natexlab{}.
\newblock \showarticletitle{Llama 2: Open foundation and fine-tuned chat models}.
\newblock \bibinfo{journal}{\emph{arXiv preprint arXiv:2307.09288}} (\bibinfo{year}{2023}).
\newblock


\bibitem[Uchida et~al\mbox{.}(2017)]%
        {uchida2017embedding}
\bibfield{author}{\bibinfo{person}{Yusuke Uchida}, \bibinfo{person}{Yuki Nagai}, \bibinfo{person}{Shigeyuki Sakazawa}, {and} \bibinfo{person}{Shin'ichi Satoh}.} \bibinfo{year}{2017}\natexlab{}.
\newblock \showarticletitle{Embedding watermarks into deep neural networks}. In \bibinfo{booktitle}{\emph{Proceedings of the 2017 ACM on international conference on multimedia retrieval}}. \bibinfo{pages}{269--277}.
\newblock


\bibitem[Wang et~al\mbox{.}(2019)]%
        {wang2019neural}
\bibfield{author}{\bibinfo{person}{Bolun Wang}, \bibinfo{person}{Yuanshun Yao}, \bibinfo{person}{Shawn Shan}, \bibinfo{person}{Huiying Li}, \bibinfo{person}{Bimal Viswanath}, \bibinfo{person}{Haitao Zheng}, {and} \bibinfo{person}{Ben~Y Zhao}.} \bibinfo{year}{2019}\natexlab{}.
\newblock \showarticletitle{Neural cleanse: Identifying and mitigating backdoor attacks in neural networks}. In \bibinfo{booktitle}{\emph{2019 IEEE symposium on security and privacy (SP)}}. IEEE, \bibinfo{pages}{707--723}.
\newblock


\bibitem[Wang and Chang(2021)]%
        {9401119}
\bibfield{author}{\bibinfo{person}{Si Wang} {and} \bibinfo{person}{Chip-Hong Chang}.} \bibinfo{year}{2021}\natexlab{}.
\newblock \showarticletitle{Fingerprinting Deep Neural Networks - a DeepFool Approach}. In \bibinfo{booktitle}{\emph{2021 IEEE International Symposium on Circuits and Systems (ISCAS)}}. \bibinfo{pages}{1--5}.
\newblock
\href{https://doi.org/10.1109/ISCAS51556.2021.9401119}{doi:\nolinkurl{10.1109/ISCAS51556.2021.9401119}}


\bibitem[Wang and Kerschbaum(2021)]%
        {wang2021riga}
\bibfield{author}{\bibinfo{person}{Tianhao Wang} {and} \bibinfo{person}{Florian Kerschbaum}.} \bibinfo{year}{2021}\natexlab{}.
\newblock \showarticletitle{Riga: Covert and robust white-box watermarking of deep neural networks}. In \bibinfo{booktitle}{\emph{Proceedings of the web conference 2021}}. \bibinfo{pages}{993--1004}.
\newblock


\bibitem[Wang et~al\mbox{.}(2025)]%
        {wang2025empowering}
\bibfield{author}{\bibinfo{person}{Xubin Wang}, \bibinfo{person}{Zhiqing Tang}, \bibinfo{person}{Jianxiong Guo}, \bibinfo{person}{Tianhui Meng}, \bibinfo{person}{Chenhao Wang}, \bibinfo{person}{Tian Wang}, {and} \bibinfo{person}{Weijia Jia}.} \bibinfo{year}{2025}\natexlab{}.
\newblock \showarticletitle{Empowering Edge Intelligence: A Comprehensive Survey on On-Device AI Models}.
\newblock \bibinfo{journal}{\emph{Comput. Surveys}} (\bibinfo{year}{2025}).
\newblock


\bibitem[Wang et~al\mbox{.}(2004)]%
        {wang2004image}
\bibfield{author}{\bibinfo{person}{Zhou Wang}, \bibinfo{person}{Alan~C Bovik}, \bibinfo{person}{Hamid~R Sheikh}, {and} \bibinfo{person}{Eero~P Simoncelli}.} \bibinfo{year}{2004}\natexlab{}.
\newblock \showarticletitle{Image quality assessment: from error visibility to structural similarity}.
\newblock \bibinfo{journal}{\emph{IEEE transactions on image processing}} \bibinfo{volume}{13}, \bibinfo{number}{4} (\bibinfo{year}{2004}), \bibinfo{pages}{600--612}.
\newblock


\bibitem[Wen et~al\mbox{.}(2023)]%
        {wen2023tree}
\bibfield{author}{\bibinfo{person}{Yuxin Wen}, \bibinfo{person}{John Kirchenbauer}, \bibinfo{person}{Jonas Geiping}, {and} \bibinfo{person}{Tom Goldstein}.} \bibinfo{year}{2023}\natexlab{}.
\newblock \showarticletitle{Tree-rings watermarks: Invisible fingerprints for diffusion images}.
\newblock \bibinfo{journal}{\emph{Advances in Neural Information Processing Systems}}  \bibinfo{volume}{36} (\bibinfo{year}{2023}), \bibinfo{pages}{58047--58063}.
\newblock


\bibitem[wikipedia.org(2025)]%
        {Stable_Diffusion}
\bibfield{author}{\bibinfo{person}{wikipedia.org}.} \bibinfo{year}{2025}\natexlab{}.
\newblock \bibinfo{title}{Stable Diffusion}.
\newblock
\urldef\tempurl%
\url{https://en.wikipedia.org/wiki/Stable_Diffusion}
\showURL{%
Retrieved March 31 2025 from \tempurl}


\bibitem[Xiong et~al\mbox{.}(2023)]%
        {xiong2023flexible}
\bibfield{author}{\bibinfo{person}{Cheng Xiong}, \bibinfo{person}{Chuan Qin}, \bibinfo{person}{Guorui Feng}, {and} \bibinfo{person}{Xinpeng Zhang}.} \bibinfo{year}{2023}\natexlab{}.
\newblock \showarticletitle{Flexible and secure watermarking for latent diffusion model}. In \bibinfo{booktitle}{\emph{Proceedings of the 31st ACM International Conference on Multimedia}}. \bibinfo{pages}{1668--1676}.
\newblock


\bibitem[Xu et~al\mbox{.}(2024)]%
        {xu2024device}
\bibfield{author}{\bibinfo{person}{Jiajun Xu}, \bibinfo{person}{Zhiyuan Li}, \bibinfo{person}{Wei Chen}, \bibinfo{person}{Qun Wang}, \bibinfo{person}{Xin Gao}, \bibinfo{person}{Qi Cai}, {and} \bibinfo{person}{Ziyuan Ling}.} \bibinfo{year}{2024}\natexlab{}.
\newblock \showarticletitle{On-device language models: A comprehensive review}.
\newblock \bibinfo{journal}{\emph{arXiv preprint arXiv:2409.00088}} (\bibinfo{year}{2024}).
\newblock


\bibitem[Yu et~al\mbox{.}(2021)]%
        {yu2021artificial}
\bibfield{author}{\bibinfo{person}{Ning Yu}, \bibinfo{person}{Vladislav Skripniuk}, \bibinfo{person}{Sahar Abdelnabi}, {and} \bibinfo{person}{Mario Fritz}.} \bibinfo{year}{2021}\natexlab{}.
\newblock \showarticletitle{Artificial fingerprinting for generative models: Rooting deepfake attribution in training data}. In \bibinfo{booktitle}{\emph{Proceedings of the IEEE/CVF International conference on computer vision}}. \bibinfo{pages}{14448--14457}.
\newblock


\bibitem[Zeng et~al\mbox{.}(2024)]%
        {zeng2024huref}
\bibfield{author}{\bibinfo{person}{Boyi Zeng}, \bibinfo{person}{Lizheng Wang}, \bibinfo{person}{Yuncong Hu}, \bibinfo{person}{Yi Xu}, \bibinfo{person}{Chenghu Zhou}, \bibinfo{person}{Xinbing Wang}, \bibinfo{person}{Yu Yu}, {and} \bibinfo{person}{Zhouhan Lin}.} \bibinfo{year}{2024}\natexlab{}.
\newblock \showarticletitle{Huref: Human-readable fingerprint for large language models}.
\newblock \bibinfo{journal}{\emph{Advances in Neural Information Processing Systems}}  \bibinfo{volume}{37} (\bibinfo{year}{2024}), \bibinfo{pages}{126332--126362}.
\newblock


\bibitem[Zeng et~al\mbox{.}(2021)]%
        {zeng2021adversarial}
\bibfield{author}{\bibinfo{person}{Yi Zeng}, \bibinfo{person}{Si Chen}, \bibinfo{person}{Won Park}, \bibinfo{person}{Zhuoqing Mao}, \bibinfo{person}{Ming Jin}, {and} \bibinfo{person}{Ruoxi Jia}.} \bibinfo{year}{2021}\natexlab{}.
\newblock \showarticletitle{Adversarial Unlearning of Backdoors via Implicit Hypergradient}. In \bibinfo{booktitle}{\emph{International Conference on Learning Representations}}.
\newblock


\bibitem[Zhang et~al\mbox{.}(2018)]%
        {zhang2018protecting}
\bibfield{author}{\bibinfo{person}{Jialong Zhang}, \bibinfo{person}{Zhongshu Gu}, \bibinfo{person}{Jiyong Jang}, \bibinfo{person}{Hui Wu}, \bibinfo{person}{Marc~Ph Stoecklin}, \bibinfo{person}{Heqing Huang}, {and} \bibinfo{person}{Ian Molloy}.} \bibinfo{year}{2018}\natexlab{}.
\newblock \showarticletitle{Protecting intellectual property of deep neural networks with watermarking}. In \bibinfo{booktitle}{\emph{Proceedings of the 2018 on Asia conference on computer and communications security}}. \bibinfo{pages}{159--172}.
\newblock


\bibitem[Zhang et~al\mbox{.}(2024)]%
        {zhang2024reefrepresentationencodingfingerprints}
\bibfield{author}{\bibinfo{person}{Jie Zhang}, \bibinfo{person}{Dongrui Liu}, \bibinfo{person}{Chen Qian}, \bibinfo{person}{Linfeng Zhang}, \bibinfo{person}{Yong Liu}, \bibinfo{person}{Yu Qiao}, {and} \bibinfo{person}{Jing Shao}.} \bibinfo{year}{2024}\natexlab{}.
\newblock \bibinfo{title}{REEF: Representation Encoding Fingerprints for Large Language Models}.
\newblock
\showeprint[arxiv]{2410.14273}~[cs.CL]
\urldef\tempurl%
\url{https://arxiv.org/abs/2410.14273}
\showURL{%
\tempurl}


\bibitem[Zhang et~al\mbox{.}(2023)]%
        {zhang2023libsteal}
\bibfield{author}{\bibinfo{person}{Jinquan Zhang}, \bibinfo{person}{Pei Wang}, {and} \bibinfo{person}{Dinghao Wu}.} \bibinfo{year}{2023}\natexlab{}.
\newblock \showarticletitle{Libsteal: Model extraction attack towards deep learning compilers by reversing dnn binary library}. In \bibinfo{booktitle}{\emph{Proceedings of the 18th International Conference on Evaluation of Novel Approaches to Software Engineering (ENASE)}}.
\newblock


\bibitem[Zhang et~al\mbox{.}(2019)]%
        {zhang2019robust}
\bibfield{author}{\bibinfo{person}{Kevin~Alex Zhang}, \bibinfo{person}{Lei Xu}, \bibinfo{person}{Alfredo Cuesta-Infante}, {and} \bibinfo{person}{Kalyan Veeramachaneni}.} \bibinfo{year}{2019}\natexlab{}.
\newblock \showarticletitle{Robust invisible video watermarking with attention}.
\newblock \bibinfo{journal}{\emph{arXiv preprint arXiv:1909.01285}} (\bibinfo{year}{2019}).
\newblock


\bibitem[Zhao et~al\mbox{.}(2023)]%
        {zhao2023recipe}
\bibfield{author}{\bibinfo{person}{Yunqing Zhao}, \bibinfo{person}{Tianyu Pang}, \bibinfo{person}{Chao Du}, \bibinfo{person}{Xiao Yang}, \bibinfo{person}{Ngai-Man Cheung}, {and} \bibinfo{person}{Min Lin}.} \bibinfo{year}{2023}\natexlab{}.
\newblock \showarticletitle{A recipe for watermarking diffusion models}.
\newblock \bibinfo{journal}{\emph{arXiv preprint arXiv:2303.10137}} (\bibinfo{year}{2023}).
\newblock


\bibitem[Zhu et~al\mbox{.}(2021)]%
        {zhu2021hermes}
\bibfield{author}{\bibinfo{person}{Yuankun Zhu}, \bibinfo{person}{Yueqiang Cheng}, \bibinfo{person}{Husheng Zhou}, {and} \bibinfo{person}{Yantao Lu}.} \bibinfo{year}{2021}\natexlab{}.
\newblock \showarticletitle{Hermes attack: Steal $\{$DNN$\}$ models with lossless inference accuracy}. In \bibinfo{booktitle}{\emph{30th USENIX Security Symposium (USENIX Security 21)}}.
\newblock


\end{thebibliography}










\end{document}